\documentclass[useAMS,usenatbib,usegraphicx]{mn2e}
\usepackage{multirow}
\usepackage{color,graphicx}
\usepackage[letterpaper]{geometry}
\usepackage{longtable,lscape}
\newcommand{\lsim}{\raisebox{-.5ex}{$\,\stackrel{\textstyle <}{\sim}\,$}}

\newcommand{\fcp}{FCP} 
\newcommand{\mnras}{MNRAS}

\newcommand{\apj}{ApJ}
\newcommand{\apjs}{ApJS}
\newcommand{\apjl}{ApJL}
\newcommand{\aj}{AJ}
\newcommand{\aap}{A\&A}

\newcommand{\araa}{ARA\&A}
\newcommand{\pasp}{PASP}
\newcommand{\icarus}{ICARUS}

\title[The low mass population of 25 Orionis]{The low mass star and sub-stellar
populations of the 25 Orionis group}
\author[J. J. Downes et al.]
{
Juan Jos\'e~Downes     $^{1,2}$\thanks{E-mail: jdownes@cida.ve,jdownes@astrosen.unam.mx},
C\'esar~Brice\~no,     $^{1,3}$\thanks{E-mail: briceno@cida.ve,cbriceno@ctio.noao.edu},
Cecilia~Mateu,         $^{1,2}$ 
Jes\'us~Hern\'andez,   $^{1}$ \newauthor
Anna Katherina~Vivas   $^{3}$
Nuria~Calvet           $^4$ 
Lee~Hartmann           $^4$ \newauthor
Monika G.~Petr-Gotzens $^{5}$ 
and Lori~Allen         $^{6}$\\
$^{1}$Centro de Investigaciones de Astronom\'{\i}a, AP 264, M\'erida 5101-A, Venezuela\\
$^{2}$Instituto de Astronom\'ia, UNAM, Ensenada, C.P. 22860, Baja California,  M\'exico\\
$^{3}$Cerro Tololo Interamerican Observatory Casilla 603, La Serena, Chile\\
$^{4}$Department of Astronomy, University of Michigan, 825 Dennison Building, 500 Church Street, Ann Arbor, MI 48109, USA\\
$^{5}$European Southern Observatory, Karl-Schwarzschild-Str. 2, 85748, Garching bei M\"unchen, Germany\\
$^{6}$National Optical Astronomy Observatoies, 950 N. Cherry Ave. Tucson, AZ 85719, USA\\
}
\begin{document}
\date{Accepted 2014 ------ --. Received 2014 ------ --; in original form ---- ------ --}
\pagerange{\pageref{firstpage}--\pageref{lastpage}} \pubyear{2014}
\maketitle
\label{firstpage}
\begin{abstract}
We present the results of a survey of the low mass star
and brown dwarf population of the 25 Orionis group.
Using optical photometry from the CIDA Deep Survey of Orion, near IR photometry
from the Visible and Infrared Survey Telescope for Astronomy and
low resolution spectroscopy obtained with Hectospec at the MMT, we
selected 1246 photometric candidates to low mass stars
and brown dwarfs with estimated masses within $0.02 \lsim M/M_\odot \lsim 0.8$
and spectroscopically confirmed a sample of 77 low mass stars as new members
of the cluster with a mean age of $\sim$7 Myr.
We have obtained a system initial mass function of the
group that can be well described by either a Kroupa power-law
function with indices $\alpha_3=-1.73\pm0.31$ and $\alpha_2=0.68\pm0.41$
in the mass ranges $0.03\leq M/M_\odot\leq 0.08$ and $0.08\leq M/M_\odot\leq0.5$ 
respectively, or a Scalo log-normal function with coefficients 
$m_c=0.21^{+0.02}_{-0.02}$ and $\sigma=0.36\pm0.03$ in the mass 
range $0.03\leq M/M_\odot\leq0.8$. 
From the analysis of the spatial distribution of this numerous candidate
sample, we have confirmed the East-West elongation of the 25 Orionis group
observed in previous works, and rule out a possible southern extension
of the group. We find that the spatial distributions of low mass stars
and brown dwarfs in 25 Orionis are statistically indistinguishable.
Finally, we found that the fraction of brown dwarfs showing IR excesses
is higher than for low mass stars, supporting the scenario in which the
evolution of circumstellar discs around the least massive objects
could be more prolonged.
\end{abstract}
%
%
%
%
%
%
\section{Introduction}\label{introduction}
Evidence collected over the past several years increasingly
supports the general picture that low mass stars (LMS) and brown dwarfs (BDs),
likely share a common formation mechanism. However, there are many open
questions that remain to be addressed. For example, given typical temperatures and densities
within molecular cloud cores \cite[e.g.][]{padoan2004}, the corresponding classical Jeans mass
can be much higher than the sub-stellar mass limit
\cite[$\sim0.072M_\odot$;][]{baraffe1998}, suggesting
that forming BDs would be difficult under such conditions.
However, we know that not only do LMS and BDs form, 
but they do so abundantly, being the most numerous
objects in the Galaxy \cite[e.g.][]{padoan2004,bastian2010}. Therefore,
their formation cannot be explained in terms of a simple-minded model of gravitational
collapse alone, which has been an important driver of theoretical and 
observational efforts looking for the physical processes and conditions 
that control LMS and BD formation.
\par
Theoretical efforts to explain the formation of LMS and BDs are 
usually classified in the following models, reviewed by 
\citet{whitworth2007}: (i) turbulent fragmentation, (ii) gravitational 
fragmentation (iii) disc fragmentation (iv) premature ejection of 
protostellar embryos and (v) photo-erosion of cores in HII regions.
These processes are not mutually exclusive \citep{whitworth2007} 
in the sense that in a typical star forming region (SFR) all of them 
could take place simultaneously, resulting in a complex scenario. The relative 
importance of each model is currently explored in terms of its
ability to reproduce observables such as the initial mass function (IMF), 
the properties of binary and multiple systems, the fraction of 
LMS and BDs showing circumstellar discs,
magnetosferic accretion and/or outflows and studying the kinematic or spatial
segregation as a function of stellar mass \cite[e.g.][]{whitworth2007}.
Each model has proven its plausibility of forming LMS and BDs,
though only a few of the available
simulations offer results directly comparable with observational data,
and several assumptions about the initial conditions
are still a matter of debate.
\par 
Observationally, several SFRs and young clusters
have been surveyed looking for their LMS and BD population
\citep[e.g.][and references therein]{luhman2012,
luhman2010,boudreault2009,bouvier2008,lodieu2007,caswell2007,luhman2007}.
Some important results are: the continuity of the mass function across 
the sub-stellar mass limit \citep[e.g.][]{bastian2010}, the existence 
of circumstellar discs \citep[e.g.][]{luhman2010}, accretion phenomena 
\citep[e.g.][]{muzerolle2000b,jayawardhana2003} and photometric variability 
\citep[e.g.][]{scholz2005} in BDs, which indicate the extension of the T-Tauri 
behaviour below the sub-stellar mass limit.
There is also some initial insight on mass segregation
\citep[e.g.][]{kumar2007,boudreault2009}, mass-dependent kinematics
\citep[e.g.][]{joergens2006} and multiplicity \citep[e.g.][]{mccarthy2004,
luhman2004a,kraus2006,luhman2009b}.
%
%
\par 
Here we present a new, deep optical/near infrared photometric survey 
of the 25 Orionis group, combined with limited follow-up spectroscopy, to
search for and characterise the LMS ($0.8\lsim M/M_\odot\lsim0.072$) 
and BD ($0.072\lsim M/M_\odot\lsim0.02$) members of this population.
The relevance of the 25 Orionis group for studies of star/disc formation and
early evolution is given by its $\sim$7 Myr old age which roughly corresponds to
the time when dissipation
of primordial circumstellar discs is expected to have finished 
\citep[$\sim$10 Myr;][]{calvet2005},
as evidenced by the decrease in disc fractions
\citep{hernandez2006,hernandez2007}. Additionally,
the study of the spatial distribution of members and/or the IMF
down to the BD regime at this age, is important to constraint the dynamical evolution
of young clusters \citep[e.g.][]{boudreault2009}.
\par
Other associations of similar age are either too spatially extended 
because of their vicinity \citep[e.g. TW Hydra, $\beta$ Pictoris, 
$\epsilon$ Cha;][]{torres2008} or too distant \citep[e.g. NGC7160, 
NGC2169, NGC2362][]{kun2008,jeffries2007,dahm2008}.
Lacerta OB1b at a distance of $\sim$400 pc and an estimated age 
within 12 to 16 Myr, is probably one of the best places to carry 
out new surveys for young LMS and BD, however, it also spans 
$\sim 70 deg^2$ \citep{chen2008}.
Another interesting region, also spanning a large area of the sky, is 
the Upper Scorpius association at a distance between 125
and 165 pc \citep{luhman2012b}, whose last age determination 
is $\sim$11 Myr old \citep{pecaut2012}.
Summarizing, the 25 Orionis group is the most numerous and
spatially dense $\sim$7 Myr old population  yet
known within $\sim$500 pc from the Sun. Its modest angular
extent ($\lsim 7 deg^2$) makes comprehensive studies of its entire population 
feasible, and being almost free of significant interstellar extinction,
it is the ideal place to study slightly more evolved young LMS and BD.
\par 
The paper is organized as follows:
In Section \ref{25ori} we summarize previous results on the
25 Orionis group.
In Section \ref{observations} we describe the observations,
data reduction, photometric candidate selection, as well 
as the spectroscopic confirmation of memberships
for a sub-sample of these candidates.
In Section \ref{extinctcontam} we estimate and analyse the visual interstellar
extinction affecting each photometric candidate and spectroscopically
confirmed member, and estimate the contamination from field
stars present in the sample of photometric candidates.
In Section \ref{analysis} we present the derivation of masses 
and ages, the analysis of the IMF, the analysis of the
spatial distribution and infrared excesses
as disc indicators and accretion signatures.
In Section \ref{conclutions} we summarize our conclusions
about the nature of the 25 Orionis group
and explore possible implications on predictions from
theoretical models for LMS and BD formation.
\section{The 25 Orionis group}\label{25ori}
%
%
\par
The 25 Orionis group was originally identified by \citet{briceno2005} 
as a $\sim 1^\circ$ radius overdensity of $\sim 20$ pre-main sequence (PMS) 
stars roughly centred on the Be star 25 Ori located at 
$\alpha_{J2000}=$05:24:44.8, $\delta_{J2000}=$+01:50:47.20. It was recognized 
on the basis of multi-epoch V and I-band photometry and follow-up optical 
spectroscopy to confirm memberships.
Their initial member sample spanned magnitudes
$14.5 \lsim V \lsim 16.5$, corresponding to masses in the range 
$0.45 \lsim M/M_\odot \lsim 0.95$\footnote{Assuming 
models from \citet{baraffe1998}, 
an age of $\sim$7 Myr and distance modulus of 7.78 mag}.
\citet{briceno2005} reported a mean age for the Orion OB1a 
sub-association, which includes the 25 Orionis 
overdensity, of $\sim 7-10$ Myr and an inner-disc frequency of
$\sim 11$ per cent, smaller than the $\sim 23$ per cent they had found in 
the younger ($\sim 3.6$ Myr) Orion OB1b sub-association. 
These results were consistent with the decrease in disc fraction
with age seen in other regions \citep{hernandez2008}, and supported 
the idea that inner discs around low-mass stars 
dissipate in less than $\sim 10$ Myr \citep{hartmann2001}.
%
%
\par 
The newly identified 25 Orionis group was also recognized
a few months later by \citet{kharchenko2005} in their extensive
survey of new galactic clusters, and listed as cluster ASCC-16.
They derive central coordinates $\alpha_{J2000}=$5:24:36, 
$\delta_{J2000}=$+01:48:00, a cluster radius=$0\fdg62$ 
and core radius=$0\fdg25$.
They estimate a distance of 460 pc ($\rm V - M_V = 8.59$),
reddening $\rm E(B-V)=0.09$, age $\sim8.5$ Myr, and a mean 
radial velocity of $0.75 \pm 8.75$ km/s. However, 
the radial velocity comes from only one star, 25 Ori itself,
and the age, distance and cluster radius are based on just a small 
number of stars earlier than K3.
%
%
\par
The low-mass population of the 25 Orionis group  was later considered by
\citet{mcgehee2006}. His variability-based survey of Orion OB1 using 
multi-epoch  photometry from the Sloan Digital Sky Survey \citep{abazajian2005}, 
within a $2\fdg5$ wide equatorial strip located roughly $0\fdg6$ south of the 
star 25 Ori. Based on the spatial distribution of his and \citet{briceno2005} 
samples, \citet{mcgehee2006} suggested that the 25 Orionis group could extend 
further south, with a total radius of $\sim 1\fdg4$, greater than previously 
reported. \citet{mcgehee2006} suggested that the 25 Orionis group is an unbound 
association and derived a Classical T-Tauri star (CTTS) fraction of $\sim 10$
per cent in what he called the 25 Orionis group.
%
%
\par
Discs around the low-mass ($0.12 \lsim M/M_\odot \lsim 1.2$) members of 25 Orionis 
were searched for and characterised by \citet{hernandez2007} in their Spitzer 
IRAC and MIPS imaging study of this group. They found that in the 25 Orionis group, 
discs around low-mass stars are less frequent ($\sim 6$ per cent) than in the younger 
($\sim 4$ Myr) Orion OB1b ($\sim 13$ per cent), and that this frequency seemed to be 
a function of the spectral type with a maximum around M0. They also determined that 
disc dissipation takes place faster in the inner region of the discs; as suggested 
by several disc models \citep[e.g.][]{weidenschilling1997,dullemond2004}.
%
%
\par
After the initial discovery and first studies, \citet{briceno2007a} 
conducted the most extensive assessment so far 
of the low-mass population ($0.3 \lsim M/M_{\odot} \lsim 1$) in the 
25 Orionis group. 
They concluded that not only is 25 Orionis a kinematically distinct 
entity, but it is also different in velocity space from the widely spread 
stellar population of Orion OB1a, within which it is located. 
These results pointed to 
the idea of the 25 Orionis group as a physical stellar group,
akin to structures like the $\sigma$ Ori cluster. From the spatial
distribution of the low-mass members they found a maximum surface 
density of $\sim 128$ stars/$deg^{2}$ with masses 
$M>0.5M_\odot$, and derived a cluster radius of $\sim$ 7 pc.
In this work they also
redetermined the fraction of Classical TTauri stars (CTTS) using 
a more numerous sample, and find
disc fractions of $\sim 6$ per cent in 25 Orionis and $\sim 13$ per cent in Ori OB1b,
confirming the factor of $2\times$ decline in the accretor fraction
seen by \citet{briceno2005}.
\par
The 25 Orionis group had been also the target for several other works:
%
%
Using the Spitzer IRAC and MIPS data,
\citet{hernandez2006} studied the debris disc fraction in stars with 
spectral types earlier than F5 in the 25 Orionis group.
%
%
\citet{biazzo2011} measured abundances for 8 T Tauri stars
with spectral types from K7 to M0 and derived an average 
metallicity $[$Fe/H$]$=-0.05$\pm$0.05.
%
%
\citet{ingleby2011}, have studied the X-ray and far ultraviolet fluxes
of 25 Orionis and determined accretion rates for a set of members
with spectral types K5 to M5.
%
%
Most recently, 25 Orionis has been a hunting ground for
groups looking for transiting planets among young solar-like 
stars \citep{vaneyken2011,neuhauser2011}.
In Table \ref{parameters} we summarize the derived
parameters we assume here for the 25 Orionis group, based on the 
aforementioned studies and the present work. 
In Figure \ref{figradec1} we show the spatial distribution of
candidates and confirmed members from previous surveys.
\begin{table*}                     
\centering
\begin{minipage}{800mm}
\caption{General properties of members and candidates for the 
25 Orionis group from the present and previous works.}
\begin{tiny}
\begin{tabular}{lccccccccc} 
\hline
Reference & Candidates & Members & Mass interval & LMS CTTS   & LMS Discs  & D     & Radius     & Age   & $\bar{A_V}$ \\
          & [N]        & [N]     &               & [per cent] & [per cent] & [pc]  & [$^\circ$] & [Myr] & [mag]       \\

\hline
\citet{briceno2005}      &  $\cdots$   & 31                  & $0.45 \lsim M/M_\odot \lsim 0.95$          &  $\sim 3$    & $\sim 3$                     & 330  & $\cdots$     & 7-10               & 0.28    \\
\citet{kharchenko2005}   &  90         & 29 \footnote{Considered by \citet{kharchenko2005} as high probability members.}         & $ M > 1.4 M_\odot $  \footnote{The most massive object predicted by \citet{baraffe1998} models at such age is $1.4 M_\odot$, corresponding to a $M_V=4.85$ while \\the fainter object in the \citet{kharchenko2005} sample is $M_V=4.05$.}                    &  $\cdots$     & $\cdots$                      & 460  & 0.62        & $\sim 8.5$         & 0.27    \\
\citet{mcgehee2006}      &  62         & $\cdots$            & $0.06 \lsim M/M_\odot \lsim 0.8$ \footnote{We interpolate into the \citet{baraffe1998} models the Bessel magnitudes obtained from the SDSS photometry using \\ \citet{davenport2006} conversions.}  &  $\sim 10$   & $\cdots$                      & 330  & 1.4         & $\cdots$            & $\cdots$ \\
\citet{hernandez2007}    &  41         & 115                 & $0.12 \lsim M/M_\odot \lsim 1.2$           &  $\cdots$     & $\sim 6$                     & 330  & $\cdots$     & $\cdots$            & $\cdots$ \\
\citet{briceno2007a}     &  $\cdots$   & 124 \footnote{A total of 47 members have kinematic information.}        & $0.3  \lsim M/M_\odot \lsim 1$             &  $\sim 6$    & $\cdots$                      & 330  & 1           & $\sim 7$           & 0.29    \\
This work                &  1246       & 77                  & $0.02 < M/M_\odot < 0.8$                         & $3.8\pm0.5$  & $4.1\pm0.4$             & 360  & 0.5         & $6.1\pm 0.8$       & 0.30    \\
\hline
\end{tabular}
\end{tiny}
\label{parameters}
\end{minipage}
\end{table*}
%
%
%
%
\section{Observations, candidate selection and membership determination}\label{observations}
\subsection{Optical photometry} \label{photometry}
Multi-epoch optical V, R, I-band, and H$\alpha$ observations across the entire
Orion OB1 association (spanning $\rm \sim 180\> deg^2$) were obtained
as part of the CVSO \citep{briceno2005}, 
being conducted since 1998 with the J\"urgen Stock 1.0/1.5 Schmidt-type 
telescope and the $8000 \times 8000$ pixel QUEST-I CCD Mosaic camera \citep{baltay2002}, 
at the National Astronomical Observatory of Venezuela. 
The instrument is composed of sixteen 
$2048\times 2048$ Loral CCD devices set in a $4 \times 4$ array
covering most of the focal plane of the telescope, and yielding
a field of view $\sim 2\fdg3$ wide in declination.
The chips are front illuminated with a pixel size of $15\mu$m, which 
corresponds to a scale of 1.02'' per pixel. Each row of four CCDs
in the N-S direction can be fitted with a different filter.
The system was optimized to operate in drift-scan mode, in which
the telescope is fixed at a constant declination and hour angle, and an 
object moves across each filter as a consequence of sidereal motion, 
resulting in a survey rate of $\sim 34$ $deg^2$ per hour per filter.
The scans for the CVSO were performed along strips centred at 
declinations $-5^\circ$, $-3^\circ$, $-1^\circ$, $+1^\circ$, $+3^\circ$ 
and $+5^\circ$ 
\citep[for more details of the survey strategy see][]{briceno2005,briceno2007a,downes2008}.
\par
Because the exposure time in the drift-scan mode observations is
fixed\footnote{The exposure time is set by how long it takes the star 
to cross the entire CCD, which in our instrument is $\sim 140$s near
the equator.}, this defines the usable magnitude range for each individual
observation. At the bright end, our data saturate around magnitude 13
in the V, R and I bands, which
corresponds to pre-main sequence (PMS) stars with masses $\sim 1 M_\odot$,
at the distance and age of the 25 Orionis group \citep{baraffe1998}.
At the faint end, the $3\sigma$ limiting magnitudes for individual scans
are $\rm V_{lim}\sim 19.7$, $\rm R_{lim}\sim 19.7$, $\rm I_{lim} \sim 19.5$,
with completeness magnitude\footnote{The magnitude for which the logarithm of the number 
of sources as a function of magnitude departs from linear behaviour} of  
$\rm V_{com}\sim 18.7$, $\rm R_{com}\sim 18.7$, $\rm I_{com} \sim 18.2$.
Again assuming \citet{baraffe1998} models, and the age and distance
of the 25 Orionis group ($\sim$7 Myr, $\sim$360 pc from 
Section \ref{agesfim}), the substellar limit is located at 
$V_{sub} \sim 19.8$, $R_{sub} \sim 18.7$ 
and $I_{sub} \sim 17$, which clearly shows that
individual scans are not sensitive enough for carrying out a 
complete study of the least massive BD in the Orion OB1 association.
\par
In order to increase the signal to noise ratio 
(SNR) and therefore the limiting and completeness magnitudes,
so that we could reach well into the substellar regime at the distance
of Orion, we coadded the individual observations
following the procedure explained in more detail in \citet{downes2008}. 
Summarizing, the process is based in the packages \texttt{Offline}
and \texttt{DQ} developed by the QUEST-I collaboration \citep{baltay2002}.
In a first step, the QUEST-I data pipeline (\texttt{Offline}) 
automatically processes every single
drift-scan considering each CCD of the mosaic as an independent device,
corrects the raw images by bias, dark current and flat field,
performs the detection of point sources, aperture photometry, 
and determines detector coordinates for each object. 
Then the software solves the world coordinate system by
computing the astrometric matrices for each CCD of the mosaic, 
based on the USNOA-2.0 astrometric catalog \citep{monet1998}. 
\par                           
After the basic processing for every individual scan 
has been done with the \texttt{Offline} package, the second step
is to coadd these reduced single observations, using the package \texttt{DQ}.
Using the transformation matrices output by \texttt{Offline}, the \texttt{DQ} package 
computes the offsets and rotations needed to place images of the 
same area of the sky, coming from different single drift-scans, in the same 
reference frame and add them pixel by pixel to produce the final 
coadded drift-scan. 
The coadded drift-scans used here
are listed in Table \ref{coaddedscans}, together with the mean coordinates, 
filter set, resulting limiting and completeness magnitudes for each 
photometric band after coadding, average FWHM of point sources, and number
of individual scans included in the coadd.
\begin{table*}
\centering
\begin{minipage}{400mm}
\caption{Observation log for the coadded drift-scans.}
\begin{tiny}
\begin{tabular}{ccccccccccccccc} 
\hline
Drift-Scan & $\delta_c$       & $\alpha_{ini}$   & $\alpha_{end}$   & Filters & FWHM \footnote{The mean FWHM is from point sources in the I-band images.} & N\footnote{The number of added single scans in the I-band.} & $V_{com}$\footnote{The limiting and completeness magnitudes were computed for a $3\sigma$ detection threshold.} & $R_{com}$ & $I_{com}$ & $V_{lim}$ & $R_{lim}$ & $I_{lim}$\\
           & $J2000 [^\circ]$ & $J2000 [^\circ]$ & $J2000 [^\circ]$ &         & $[^{\prime\prime}]$                                                              &                                                             &                                                                                                                &           &           &           &           &          \\
\hline
Single          & +1, +3 & 65.00 & 99.00 & All filters  & 2.5 &  1  & 18.7 & 19.0  & 18.5 & 20.5 & 20.5  & 20.0 \\ 
Coadds 011-012  & +1     & 76.14 & 86.30 & VIIV         & 2.6 & 28  & 20.5 & \dots & 20.3 & 22.3 & \dots & 21.8 \\
Coadd  013      & +1     & 77.11 & 82.56 & VIRI         & 2.8 & 29  & 20.5 & 20.8  & 20.3 & 22.3 & 22.3  & 21.8 \\
Coadd  940      & +1     & 67.00 & 98.22 & RH$\alpha$IV & 2.9 & 13  & 20.1 & 20.4  & 19.9 & 21.9 & 21.9  & 21.4 \\
Coadd  950      & +3     & 71.15 & 89.41 & VRIV         & 2.6 & 10  & 19.9 & 20.2  & 19.7 & 21.7 & 21.7  & 21.2 \\
\hline
\end{tabular}
\end{tiny}
\label{coaddedscans}
\end{minipage}
\end{table*}
\par
In a third step, photometry and astrometry on the resulting coadded images are 
performed using an IRAF\footnote{IRAF is distributed by the
National Optical Astronomy Observatories, which are operated by the Association
of Universities for Research in Astronomy, Inc., under cooperative agreement
with the National Science Foundation.}
and IMWCS-based pipeline\footnote{IMWCS is the Image World Coordinate Setting 
Program described in \citet{mink2006}}. The maximum
number of detections at the $3\sigma$ threshold in the I-band 
images, corresponds to a mean spatial density of  $\sim 1$
point source per $\sim 340$ arcsec square, resulting in a mean
distance between nearest sources of about $\sim 22"$. Because of this,
source crowding is not a problem, and we can safely
perform aperture photometry; we used an aperture 
radius equal to the average FWHM of the coadded images ($\sim 3"$; Table \ref{coaddedscans}). 
\par
The zero point calibration of the instrumental photometry in the Cousins system
for each coadded drift-scan was done by normalizing the instrumental magnitudes 
to a master catalogue of $\sim10^5$ non-variable stars\footnote{Since we used
individual scans spanning a time baseline of 7 years, we could determine
which stars did not show hints of variability above the 0.02 mag level, and
use these as secondary standards for our photometric calibration. Because of
this, and the large number of stars used, we obtained a very robust calibration.} 
with calibrated magnitudes. This large set of secondary standards was produced
from a set of single drift-scans obtained with the same instrument
and filters under photometric conditions,
calibrated with \citet{landolt1992} standard star fields \citep{mateu2012}. 
The calibration has an RMS of 0.021, 0.037 and 0.042 magnitudes in the 
$V$, $R$ and $I$ bands respectively. 
The mean positional uncertainty as inferred from the 
offsets in coordinates between the optical bands is 0.4\arcsec, with a 
standard deviation of 0.6\arcsec, similar to what we find 
for the offsets with respect to the infrared datasets we have used 
in this work (\S\ref{irphotometry}).
Therefore, the photometric and astrometric calibration offer adequate 
accuracy for a reliable selection of LMS and BD candidates.
\par
The catalogues with calibrated photometry for each photometric 
band from different coadded drift-scans, were finally compiled  
into a single deep catalog for each band,
extracting repetitions and keeping, in such cases, always the 
measurement showing the highest SNR\footnote{All positional
matching between object catalogs throughout this paper was 
done using routines based on the the Starlink Tables 
Infrastructure Library Tool Set \citep[STILTS,][]{taylor2006}
available at http://www.starlink.ac.uk/stilts/}.
We matched our optical source lists
using a search radius of 3\arcsec, always keeping the object with the minimum
positional difference. We will refer to this collection
of four deep catalogs (in the V, R, I passbands and in a 
narrow band - 10 nm wide - H$\alpha$ filter),  
as the CIDA Deep Survey of Orion (CDSO, Downes et al. 2014 in preparation).  
The sources in the CDSO 
span the region $\rm 65^\circ \le \alpha_{J2000} \le 100^\circ$ 
and  $-6^\circ \le \delta_{J2000} \le +6^\circ$, with positions and
photometry for 8.6 million objects in the I-band, 8.1 million objects
in the R-band, 6.9 million objects in the V-band, and 2.4 million
objects in the H$\alpha$ filter. 
\par
In this contribution we focus on the LMS and BD populations of the 25
Orionis group, therefore we limit our analysis and discussion to drift-scans centred at
declinations $+1^\circ$ and $+3^\circ$, in particular
within the region $0\fdg35 \lsim \delta \lsim 3\fdg35$
and $79\fdg7 \lsim \alpha \lsim 82\fdg7$
shown in Figures \ref{figradec2},
which encompasses the area expected to be spanned by the 25 Orionis
group and its surroundings.
In this area the I-band catalog has a limiting
$\rm I_{lim} \sim 22$ and completeness $\rm I_{com} \sim 19.6$,
which is enough to detect objects down to masses $\rm \sim 30\> M_{Jup}$,
assuming \citet{baraffe1998} models for the assumed age and distance of 25 Orionis.
The full analysis of the entire CDSO will be presented in a forthcoming article.
\begin{figure}
\includegraphics[width=80mm]{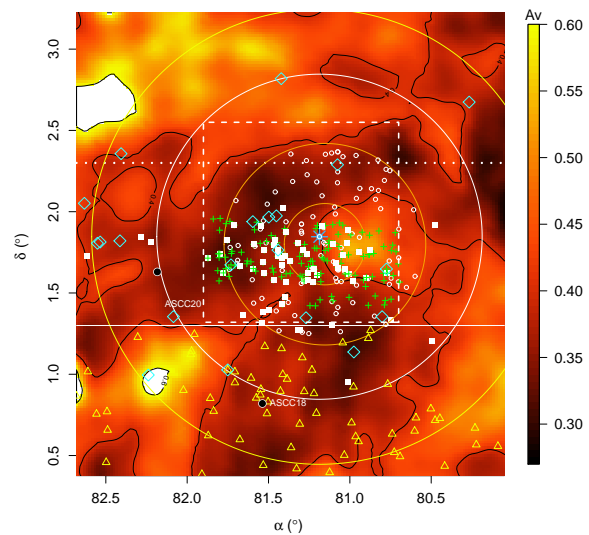}
\caption{Spatial distribution of photometric candidates and confirmed members 
of the 25 Orionis group from previous studies. Solid squares: LMS confirmed 
as members from \citet{briceno2005,briceno2007a}. The dotted horizontal line 
indicates the northern limit of the survey from \citet{briceno2005}. Small 
circles: Photometric and astrometric candidates from \citet{kharchenko2005}. 
Crosses: members and candidates from \citet{hernandez2006}. Triangles: Photometric 
candidates from \citet{mcgehee2006}. The solid horizontal line indicates the 
northern limit of the survey of \citet{mcgehee2006}. Diamonds: Eclipsing systems 
and/or CTTS candidates from \citet{vaneyken2011}. Large circles indicate 
from the centre outward: the core and corona cluster radii 
($0\fdg25$ and $0\fdg62$ respectively) computed by \citet{kharchenko2005}, 
the radius suggested by \citet{briceno2005} ($\sim1^\circ$) and the radius 
proposed by \citet{mcgehee2006} ($\sim1\fdg4$); the dashed square indicates 
the spatial limits of the IRAC survey. 
Background: Map of the extinction A$_V$ from \citet{schlegel1998} with scale on the right.
The big central asterisk represents the 25 Ori star and the black dots other
clusters from \citet{kharchenko2005}.}
\label{figradec1}
\end{figure}
\begin{figure}
\includegraphics[width=80mm]{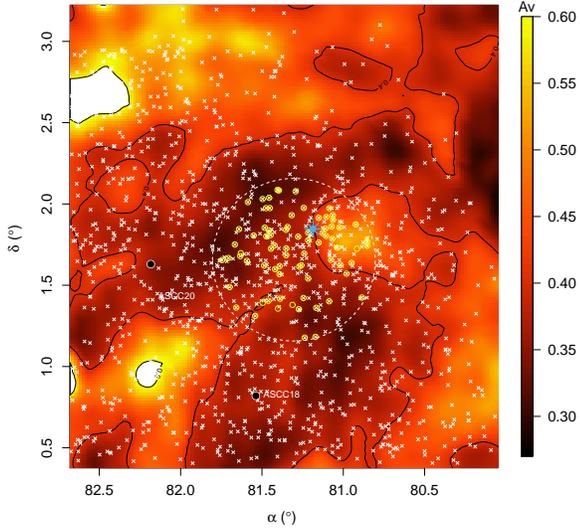}
\caption{Spatial distribution of LMS and BD of the 25 Orionis group 
from the present work. Crosses: LMS and BD photometric candidates 
(\S\ref{candidates}). Circles: LMS confirmed as members 
(\S\ref{spec_clasif}). The dashed circle indicates the field observed 
with Hectospec (\S\ref{spectroscopy}).
Other elements are as in Figure \ref{figradec1}.}
\label{figradec2}
\end{figure}
\subsection{Infrared photometry}\label{irphotometry}
\subsubsection{Near Infrared Data}
As will be explained in Section \ref{candidates}, our technique of
searching for young BDs and LMS in the extended off-cloud, 
low-extinction regions of the Orion OB1 association
hinges on the combination of deep optical and near-infrared
photometry. Ever since its release, the Two Micron All-Sky Survey 
\citep[2MASS][]{skrutskie2006}
has been the tool of choice for all large scale studies
that required access to near infrared photometry for large numbers of
sources over  extended areas of the sky, with uniform spatial coverage.
However, with limiting magnitudes of $\rm J_{lim} \sim 17.4$, 
$\rm H_{lim} \sim 16.7$, $\rm Ks_{lim} \sim 16.2$ and completeness 
magnitudes $\rm J_{com} \sim 16.6$, $\rm H_{com} \sim 15.8$, $\rm Ks_{com} \sim 15.4$,
the 2MASS catalog is shallow compared to the deep CDSO data,
imposing restrictions on our search for BD in Orion,
in particular for the faintest, lowest mass objects. 
Nevertheless, being at the time of our initial candidate selection the only
spatially complete near-IR dataset available, we cross matched our I-band 
CDSO catalog with 2MASS using a search radius of 3\arcsec,
resulting in a master catalog with IJHKs photometry for each source. 
With this dataset we created the optical-near-IR colour-magnitude
diagrams used to select LMS and BD candidates (see \S \ref{candidates}) 
for follow-up spectroscopy.
\par
Recently, a new, much deeper near-IR survey has become available.
During 2009 a new dedicated 4m survey telescope, the Visible and Infrared Survey 
Telescope for Astronomy (VISTA), located at ESO's Paranal Observatory, was 
commissioned by the VISTA consortium \citep{emerson2004,emerson2010}.
It is equipped with a 67$\times$10$^{6}$ pixels camera with a scale
of 0.34\arcsec/pix and a field of view of $1\fdg65$ in diameter.
For the Galactic Science Verification of VISTA, a $\rm \sim 30\> deg^2$ area
of the Orion OB1 association, which included the Orion Belt region, part of the Orion
A cloud, the 25 Orionis and $\sigma$ Ori clusters, was imaged in the Z, Y, J, H and Ks filters,
during October 16 to November 2, 2009, down to $5\sigma$ limiting magnitudes
$\rm Z\sim 22.5$, $\rm Y\sim 21.2$, $\rm J\sim 20.4$, $\rm H \sim 19.4$, $\rm Ks \sim 18.6$
\citep{petrgotzens2011}.
The data reduction was performed using a dedicated pipeline, developed within the VISTA
Data Flow System, and run by the Cambridge Astronomy Survey Unit in the UK. 
The pipeline delivers science-ready stacked images and mosaics, as well as photometrically 
calibrated source catalogs\footnote{More details on the VISTA data processing can be 
found at http://casu.ast.cam.ac.uk/surveys-projects/vista/technical.}.
The final, band-merged Orion OB1 VISTA catalog contains $\sim$3 million sources.
The VISTA photometric calibration is based on (but different from) 2MASS.
Instrumental magnitudes were converted into apparent calibrated magnitudes 
on the VISTA system and calibrated via colour equations. This also 
included the Z and Y bands, following the procedure described in \citet{hodgkin2009}.
The point sources detected in the I-band CDSO survey catalog were cross-matched
with the VISTA Orion OB1 master catalog, using a 3\arcsec
match radius. The resulting IZYJHKs catalog
was then matched with our V, R and H$\alpha$ catalogs to add additional
measurements in these bands for those objects which had them; however,
the VRH$\alpha$ data were not used in the process of selecting LMS and BD candidates.
Because VISTA goes much deeper than 2MASS, and since for objects in common the 
VISTA and 2MASS magnitudes agree within the uncertainties, we adopted the CDSO-VISTA dataset as
our final optical-near IR photometry catalog.
For objects showing $J<12$, $H<12$ or $Ks<11$ in the VISTA catalogue we adopted 
J, H and Ks magnitudes from 2MASS in order to avoid possible saturation effects 
in VISTA data.
\subsubsection{Mid-infrared Data}
In order to identify disc-bearing sources among selected candidate LMS and BDs,
we matched our multiband VRH$\alpha$IZYJHKs catalog with our existing Spitzer observations
within the 25 Orionis region obtained as part of GO-3437 \citep[]{hernandez2007} and 
GO-50360 (Brice\~no et al. 2014, in preparation) and also with the recently 
released Wide Infrared Survey Explorer \citep[WISE - ][]{wright2010} database.
First, a sub-sample of the photometric candidates and of the 
spectroscopically confirmed members (Sections \ref{candidates} and \ref{spec_clasif}), 
were identified in the object catalogs extracted from our IRAC images.
The image reduction, point source detection and photometry
for the datasets from \citet{hernandez2007} and 
Brice\~no et al. (2014, in preparation) are analogous 
and were described in \citet{hernandez2007}. 
The IRAC observations were performed in the pass-bands
centred at 3.6 $\mu$m, 4.5 $\mu$m, 5.8 $\mu$m and 8.0 $\mu$m
with the spatial coverage shown in Figure \ref{figradec1}.
As a second step, the photometric candidates and spectroscopically 
confirmed members were identified in the WISE database which offers 
measurements in four pass-bands, centred at 3.4 $\mu$m, 4.6 $\mu$m, 
12  $\mu$m and 22 $\mu$m. 
Compared to the IRAC observations, WISE is a shallow survey
with worse angular resolution, but its spatial coverage is uniform, 
providing also data for the surroundings of 25 Orionis, where we did 
not have Spitzer data.  
The limiting and completeness magnitudes for each
VISTA, IRAC and WISE photometric bands are summarized in 
Table \ref{limitingIR} together with the number of photometric
candidates and confirmed members from this work with
available photometry from these IR surveys.
\begin{table*}
\centering
\begin{minipage}{140mm}
\caption{Limiting and completeness magnitudes and the number
of candidates and members from VISTA, IRAC and WISE surveys.}
\begin{tabular}{cccccc}
\hline
Survey & Band & Completeness    & Limiting & Candidates \footnote{VISTA and WISE surveys cover the entire region while IRAC survey covers the area shown in Figure \ref{figradec1}.} & Members \\
\hline
VISTA           & Z             & 22.5 & 23.5 & 1246 & 77 \\
                & Y             & 20.8 & 22.5 & 1246 & 77 \\
                & J             & 20.2 & 21.7 & 1246 & 77 \\
                & H             & 19.2 & 20.5 & 1246 & 77 \\
                & Ks            & 18.4 & 19.6 & 1246 & 77 \\
\hline
IRAC            & $3.6\mu m$    & 17.2 & 19.5 &  369 & 72 \\
                & $4.5\mu m$    & 16.7 & 19.0 &  388 & 77 \\
                & $5.8\mu m$    & 15.7 & 18.5 &  365 & 70 \\
                & $8.0\mu m$    & 14.7 & 17.5 &  358 & 77 \\
\hline
WISE            & $3.4\mu m$    & 16.2 & 17.2 & 1038 & 77 \\
                & $4.6\mu m$    & 16.0 & 17.0 & 1038 & 77 \\
                & $12 \mu m$    & 12.8 & 13.2 & 1038 & 77 \\
                & $22 \mu m$    & 9.1  & 9.5  & 1038 & 77 \\
\hline
\end{tabular}
\label{limitingIR}
\end{minipage}
\end{table*}
\subsection{Selection of photometric candidates}\label{candidates}
\par
The selection of photometric candidates was performed 
according to their position in colour-magnitude diagrams
which combine optical and near infrared magnitudes. This method 
has been extensively used and has proved highly successful 
in identifying LMS and BDs
in young clusters and star-forming regions
\citep[e.g][]{briceno2002,luhman2003b,downes2008}.
\par
\begin{figure}
\includegraphics[width=80mm]{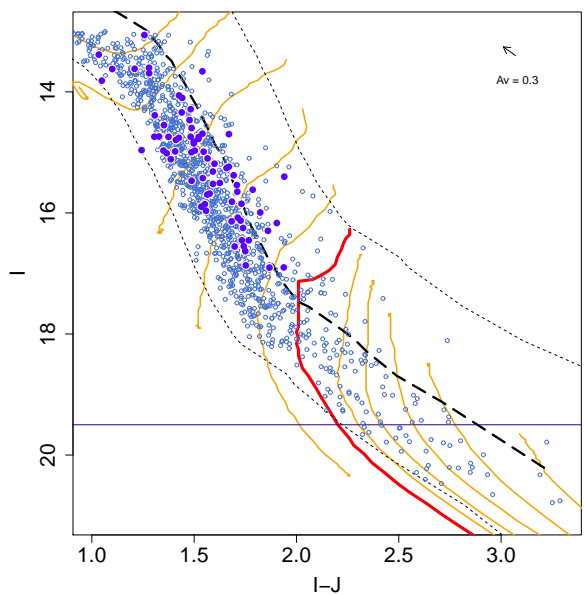}
\caption{Color-magnitude diagram used as part of the candidate 
selection, including the objects selected as candidates 
(empty circles) and those confirmed as new members (big dots).
The thick horizontal line indicates the completeness limit.
The solid curves indicates evolutionary tracks for 0.8 M$_\odot$, 
0.7 M$_\odot$, 0.5 M$_\odot$, 0.4 M$_\odot$, 0.1 M$_\odot$, 0.06 M$_\odot$, 
0.05 M$_\odot$, 0.03 M$_\odot$, 0.02 M$_\odot$, 0.01 M$_\odot$ (thin) and 0.072 M$_\odot$ (thick).
The thick dashed line indicates the isochrone corresponding to 7 Myr
and the thin doted lines enclose the membership locus.
Theoretical isochrones and evolutionary tracks
are from \citet{baraffe1998} models, are ploted without
extinction correction and assuming $m-M=7.78$ magnitudes.
The dereddening vector for the mean extinction throughout the region 
($A_V=0.3$) is also indicated.}
\label{figcm1}
\end{figure}

We selected as candidates those objects located
in what we call the \emph{membership locus}, simultaneously 
in the I vs I-J, I vs I-H
and I vs I-Ks colour-magnitude diagrams.
These loci were defined as regions centred on the isochrone 
corresponding to the mean age from the \citet{baraffe1998} models
($\sim$7 Myr) reported by \citet{briceno2005,briceno2007a}
and a distance of $\sim$360 pc (Brice\~no et al. 2014, in preparation).
The width of the loci were computed considering 
the 1 $\sigma$ uncertainties in age and distance, the mean
interstellar extinction throughout the region, unresolved 
binarity, the 1 $\sigma$ photometric uncertainties and the 
expected mean intrinsic photometric variability. The resulting 
locus for the I vs. I-J diagram is shown in Figure \ref{figcm1}.
An object was considered as a photometric candidate if 
it was located within the membership locus in all three diagrams.
Using the CDSO-VISTA catalog, a total of 1246 photometric candidates 
were selected inside the region showed in Figure \ref{figradec2}, 
spanning a $\sim$ $2\fdg5$ $\times$ $\sim$ $2\fdg5$ area covering 
the previously known population of the 25 Orionis group and 
its surroundings. 
\par
The estimated masses of the full candidate sample range from
$\sim 0.8 M_\odot$ down to $\sim 0.02 M_\odot$ according to the models of 
\citet{baraffe1998}, with an age of $\sim$7 Myr and distance modulus of 
7.78 mag.
Due to the sensitivity and saturation of the I-band observations,
the sample is complete from $\sim 0.8 M_\odot$ ($I\sim13$) down to
$\sim 0.03 M_\odot$ ($I\sim19.6$).
In Section \ref{agesfim} we present the estimation of ages and
masses for each candidate.
\par
We emphasize that colour-magnitude diagrams constructed
combining optical and near infrared magnitudes
yield the best selection of photometric PMS stellar candidates 
and young sub-stellar objects mainly because of two reasons: 
first, diagrams based on optical 
data alone (V, R and I-band magnitudes) suffer from an important 
completeness bias because BDs at this age and distance are red and 
very faint, so they drop out quickly in the bluer bands. 
Second, although our selection is not exempt from 
contamination by field stars, candidate selection performed
in diagrams that include only near infrared data strongly 
increase the contamination by field dwarfs and by extragalactic 
sources, because of the limited leverage in colour; 
in these diagrams the isochrones are almost 
vertical and piled up, therefore, there is hardly any separation between 
the PMS locus and the field population. The contamination in our candidate 
sample by field stars and extragalactic sources is discussed 
in Section \ref{contamination}.
The photometric variability in young LMS and BDs affects the 
optical and near-IR bands in a different way. Because our I-band
data is the result from a coadd of images obtained in a period 
of 15 months, we considered that as the mean I-band magnitudes. 
The reported amplitudes of the variations in J, H and Ks-bands
for LMS and BDs are within 0.2 to 0.5 mag \cite[e.g.][]{scholz2009b}
We add to the width of the membership locus the corresponding 
change in colors that result from the maximum variation of 0.5 
magnitudes in the near-IR bands.
\subsection{Optical spectroscopy} \label{spectroscopy}
Even with a refined selection technique, photometric candidate samples
suffer from some degree of contamination by non-members like foreground field dwarfs
and also by background giant stars and extragalactic objects.
Therefore, spectroscopic follow up is necessary to unambiguously confirm membership.
Moreover, spectra also provide crucial information necessary to
derive fundamental properties, like the object's effective temperature,
$\rm A_V$, and hence its luminosity, whether it is accreting from a circumstellar disc,
if it has a jet, and so on.
Here we have extended in number and to much lower masses the member
list discussed by \citet{briceno2005,briceno2007a}, 
by obtaining low resolution (R$\sim$1000) optical spectra of a sample 
of 90 photometric candidates of the 25 Orionis group. 
\par
The spectra of the sample were obtained during the night of October 22 
2007, during queue mode observations 
with the Hectospec multi-fiber spectrograph on the 
6.5 MMT telescope \citep{fabricant2005}. We observed 
a single field centred at $\alpha_{J2000}=$05:25:13, 
$\delta_{J2000}=$+01:39:25
(Figure \ref{figradec2}), using one single fiber configuration 
for which we obtained five consecutive exposures, each 1800 s long.  
We used the 270 ${\rm groove \, mm^{-1}}$ grating, providing a spectral 
resolution of R$\sim$1000 with a spectral coverage from 3700{ \AA}
to 9150{ \AA}. Because of the poor seeing conditions ($\sim2.2^{\prime\prime}$) 
during the integrations and the diameter of the Hectospec
fibers ($\sim1.5^{\prime\prime}$), only the brightest candidates 
($I\lsim17$) could be observed with reasonable SNR ($\sim15$), allowing 
for an appropriate spectral classification.
\par
The 25 Orionis group is placed far enough from the
Orion molecular clouds, so that problems due to the high 
sky background caused by nebulosity are not present.
All the Hectospec spectra were processed, extracted, corrected for sky lines,
 and wavelength calibrated by 
S. Tokarz at the CfA Telescope Data Center, using customized IRAF routines 
and scripts developed by the Hectospec team \citep[see][]{fabricant2005}. 
The five individual 1800~s exposures were coadded into a single, deep
observation, with a total integration time of 9000~s.
Light sources inside the Hectospec fiber positioner, contaminate $\sim 1/3$ 
of the fibers at wavelengths longer than $\sim 8500$ {\AA } (N. Caldwell, 
private communication), therefore, during the analysis we ignored data at 
these wavelengths.
\par
\begin{figure}
\includegraphics[width=80mm]{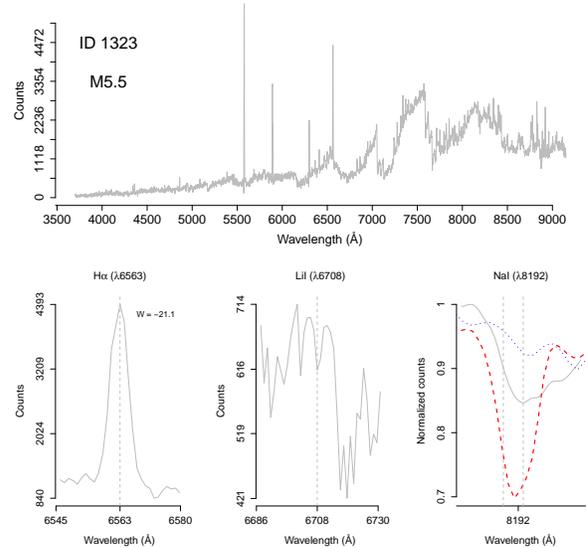}
\caption{Spectra of a LMS confirmed in this work 
as a new member of the 25 Orionis group.
The top panel shows the complete spectrum obtained by Hectospec
and the lower panels
the H$\alpha$ emission line, the LiI $\lambda6708$ 
absorption line and the NaI $\lambda 8183,8195$ absorption doublet.
Labels indicate the member ID, the spectral type and the equivalent 
width of H$\alpha$. Vertical dashed lines in the lower panels 
indicate the central wavelength of the H$\alpha$, LiI $\lambda6708$ 
and NaI $\lambda 8183,8195$.
In the panel for the NaI $\lambda 8183,8195$ absorption doublet we
also show the corresponding spectra of a field dwarf (dashed red line;
\citet{kirkpatrick1999}), and a young dwarf from the Chamaeleon I 
region ($\sim$2 $Myr$; dotted line; \citet{luhman2005}) of the same 
spectral type and smoothed to a same spectral resolution of 16 {\AA}.
The spectra for the entire sample of new members (see Table \ref{alldata1}) 
are available in the electronic edition.}
\label{spectravlms}
\end{figure}

Of the 1246 photometric candidates, 277 are placed within our Hectospec field. 
Because of limitations with the positioning of fibers and the poor seeing 
conditions we observed a total of 90 candidates out of which 77 were confirmed 
as new members of the 25 Orionis group for a $\sim$86 per cent efficiency.
The remaining 13 candidates were confirmed as field stars. An example of the 
spectra is shown in Figure \ref{spectravlms}.
\subsection{Spectral classification and membership diagnosis}\label{spec_clasif}
We performed the spectral classification by
comparing the equivalent widths of sixteen
spectral features typical of M type stars, such as TiO and VO bands,
\citep[see Table 3 of][]{downes2008}, 
with their corresponding values in 
a library of standard spectra of field dwarfs from 
\citet{kirkpatrick1999}, and following the semi-automated 
scheme of \citet{hernandez2004}\footnote{This 
procedure was done using the code SPTCLASS, available at 
http://www.astro.lsa.umich.edu/$\sim$hernandj/SPTclass/sptclass.html}.
The differences in surface gravity between the young targets and the 
field dwarfs used as standards did not affect significantly the 
spectral classification for objects earlier than M6. We estimate 
the typical uncertainty resulting from our classification to be 
$\pm$0.5 spectral subclass.
\par
We have used several criteria to assign 25 Orionis group membership 
to each  spectroscopically observed candidate. We start by considering the 
detection of H$\alpha$ in emission. Strong H$\alpha$ 
in emission is a common feature of chromospherically active young objects, 
and of objects accreting from a circumstellar disc for the largest equivalent 
widths \cite[e.g.][]{muzerolle2005}. 
Figure \ref{spectravlms} shows the H$\alpha$ emission
line profile and its corresponding equivalent width for a confirmed member.
At the age of the 25 Orionis group ($\sim$7 Myr), most stars have
stopped accreting (at least at significant levels), such that strong 
H$\alpha$ emission is not necessarily expected \citep{briceno2007a},
and many members could exhibit equivalent widths similar 
to those measured in active dMe stars. Therefore, as an additional membership indicator 
we looked at the NaI $\lambda 8183,8195$ absorption doublet. 
The NaI $\lambda 8183,8195$ feature is
sensitive to the surface gravity of the star, being very strong (deep)
in old field dwarf stars, very weak in late type M giants, and
intermediate in late type PMS stars and young BDs
which are still contracting down to the main sequence 
or asymptotically approaching the main sequence in the case of BDs
\citep{luhman2003a}.
The NaI $\lambda 8183,8195$ absorption doublet of one of the LMS 
confirmed as member, superimposed on spectra of field dwarfs from
\citet{kirkpatrick1999}, and of young M dwarfs of the same spectra
type \citep[from][]{luhman2003b} are shown in Figure \ref{spectravlms}.
Given the typical I-J, I-H and I-Ks colours of the LMS and BD candidates considered here and
the low visual extinction along the line of sight towards the 25 Orionis group,
no reddened late giant stars are expected to contaminate our candidate sample,
as we show in Section \ref{contamination}.
\par
In addition to the above criteria, 
the presence of the Li I $\lambda 6708$ line in absorption
was also considered. This spectral feature is a well-known 
indicator of youth in late K and M-type objects 
\citep{strom1989,briceno1998,briceno2001}, up to a spectral type $\sim$ M6.5,
which at an age of $\sim$10 Myr corresponds to the mass
limit for Li fusion \citep[$M \sim 0.06M_\odot$][]{basri1998}.
However, because of the low spectral resolution and SNR in most 
of our spectra, we could detect the LiI $\lambda 6708$ line
in only $\sim60$ per cent of the confirmed members. 
Objects showing Li I $\lambda 6708$ line in absorption were
selected as members; those without a detection were not
discarded as members unless they also failed the 
Na I criteria explained above.
\par
Summarizing, photometric candidates were classified as 
members if they showed H$\alpha$ in emission
\emph{and} NaI $\lambda 8183,8195$ in absorption weaker than a field dwarf
of the same spectral type. The detection of 
the Li I $\lambda 6708$ line in absorption was only considered as an 
additional youth indicator, in light of
the bias imposed by the low spectral resolution
and the SNR in most spectra.
Additionally, as a sanity check, we looked at the $A_V$ of each star,
computed from the observed colours and from the intrinsic colours we
obtained from the spectral types and the relations from \citet{luhman2003b}, 
and noted whether it was consistent with the average $A_V$ for the 25 Orionis 
group \citep{briceno2007a}.
\par
Clearly, the spectral features we have considered above 
are youth indicators but do not, by themselves, 
guarantee membership to the 25 Orionis group.
The reason is that these same criteria will select young members
that belong to other regions of the Orion OB1 association, 
placed behind the 25 Orionis group. Indeed, we
know that 25 Orionis is located in the extended OB1a sub-association,
and the much younger OB1b region is a few degrees away.
However, as we argue in Section \ref{contamination} the 
expected number of such possible young contaminants is low 
and does not affect the general analysis presented here.
Additionally, the Hectospec field is centred in
the 25 Orionis spatial overdensity we explain in 
Section \ref{spatial}, increasing the probability
of these objects being real members of 25 Orionis group.
\par
\begin{figure}
\includegraphics[width=80mm]{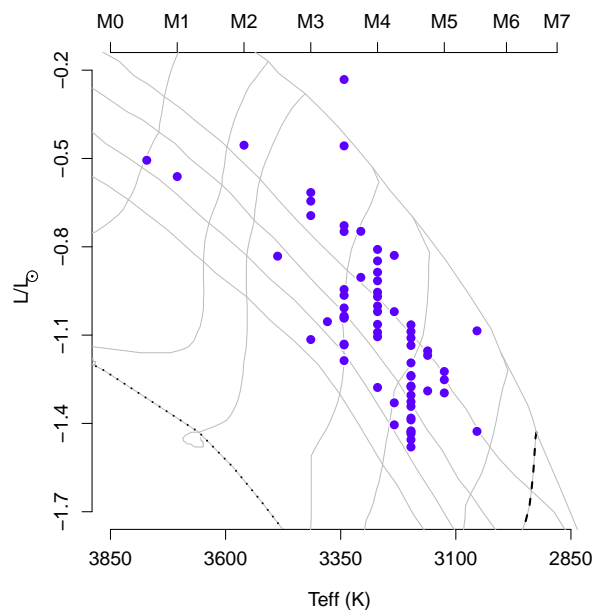}
\caption{H-R diagram for the new members. 
The isochrones for 1, 3.2, 4.6, 8.5, 11.4 and 100 (dotted) Myr and the evolutionary 
tracks for 0.8 M$_\odot$, 0.6 M$_\odot$, 0.5 M$_\odot$, 0.3 M$_\odot$, 0.2 M$_\odot$, 0.1 M$_\odot$ 
and 0.072 M$_\odot$ (dashed) from the \citet{baraffe1998} models are indicated.
The top axis indicates the corresponding spectral types from \citet{luhman2003b}.}
\label{memberscmhr}
\end{figure}

We confirmed 77 new members of the group whose photometric 
information and resulting spectral types are listed 
in Tables \ref{alldata1} and \ref{alldata2}. 
In Figure \ref{memberscmhr} we show the H-R diagram with 
the new members.
\begin{table*}
\centering
\begin{minipage}{170mm}
\caption{Photometric catalog of the new spectroscopically confirmed
LMS of the 25 Orionis group. The table is published in its entirety in the electronic edition.}
\begin{scriptsize}
\begin{tabular}{lllcccccccccccc}
\hline
ID      & $\alpha$ J2000  & $\delta$ J2000 & $V_C$ & $R_C$ & $I_C$ & Z     & Y     & J      & H     & Ks    & 3.6$\mu$m &  4.5$\mu$m & 5.8$\mu$m & 8.0$\mu$m \\ 
        & [$^\circ$]      & [$^\circ$]     &       &       &       &       &       &        &       &       &           &            &           &           \\ 
\hline                                                                                     
   1232 & 81.2582739      & 1.6224949      & 16.25 & 15.18 & 13.69 & 13.35 & 12.95 & 12.36  & 11.71 & 11.50 & 11.45 & 11.24 & 11.38 & 11.18 \\ 
   1242 & 81.0372844      & 1.7311253      & 16.11 & 15.09 & 13.66 & 13.18 & 12.68 & 12.07  & 11.57 & 11.39 & 11.17 & 11.11 & 11.07 & 11.10 \\ 
   1245 & 81.0632586      & 1.9016645      & 20.61 & 19.24 & 16.90 & 16.32 & 15.56 & 14.98  & 14.51 & 14.16 & 13.77 & 13.72 & 13.67 & 13.65 \\ 
   1246 & 81.2778866      & 1.7987624      & 19.16 & 17.83 & 15.65 & 15.12 & 14.46 & 13.89  & 13.40 & 13.11 & 12.69 & 12.64 & 12.57 & 12.54 \\ 
   1247 & 81.0647164      & 1.9310005      & 17.22 & 16.03 & 14.28 & 13.85 & 13.28 & 12.75  & 12.25 & 11.96 &       &       &       &       \\ 
   1248 & 81.2432939      & 1.9759851      & 17.45 & 16.38 & 14.73 & 14.33 & 13.86 & 13.36  & 12.77 & 12.56 &       & 12.22 &       & 12.17 \\ 
   1252 & 81.270781       & 1.6148316      & 16.92 & 15.81 & 14.1  & 13.64 & 13.17 & 12.61  & 12.04 & 11.81 & 11.49 & 11.49 & 11.48 & 11.47 \\ 
   1253 & 81.1664269      & 1.3613049      & 15.29 & 14.36 & 13.38 & 12.96 & 12.63 & 12.30  & 11.60 & 11.42 & 11.29 &       & 11.25 &       \\ 
   1258 & 81.2160937      & 1.541317       &       & 18.58 & 16.29 & 15.72 & 14.98 & 14.38  & 13.91 & 13.56 & 13.17 & 13.10 & 13.11 & 13.14 \\ 
   1261 & 81.0869439      & 1.4013985      & 15.69 & 14.74 & 13.62 & 13.32 & 13.00 & 12.47  & 11.77 & 11.65 & 11.47 &       & 11.40 &       \\ 
\hline
\end{tabular}
\end{scriptsize}
\label{alldata1}
\end{minipage}
\end{table*}
\begin{table*}
\centering
\begin{minipage}{170mm}
\caption{Spectroscopic catalog of the new confirmed members
of the 25 Orionis group. The table is published in its entirety in the electronic edition.
}
\begin{scriptsize}
\begin{tabular}{rcccccc}
\hline
ID & ST   & W[H$\alpha$] & LiI \footnote{The flags -1, 0 and 1 indicate respectivelly non detection, uncertain detection and certain detection.} & Av  \\
   &      &[$\AA$]      &    & [mag]\\
\hline
1278           &       M4.5    &       -5.8    &       0       &       0 \\
1273           &       M5      &       -13     &       1       &       0.12 \\
1262           &       M4.5    &       -8.4    &       1       &       0 \\
1277           &       M3.5    &       -4.3    &       -1      &       0 \\
1284           &       M4.5    &       -9.9    &       1       &       0 \\
1269           &       M4.5    &       -16.7   &       1       &       0.46 \\
1263           &       M2.5    &       -5      &       1       &       0.54 \\
1266           &       M4      &       -8.9    &       1       &       0.06 \\
1279           &       M3.5    &       -8.9    &       0       &       0.21 \\
1242           &       M3.5    &       -6.6    &       1       &       0.79 \\
\hline
\end{tabular}
\end{scriptsize}
\label{alldata2}
\end{minipage}
\end{table*}
\section{Interstellar extinction and contamination by field stars}\label{extinctcontam}
One of the problems encountered when carrying out photometric 
surveys over large areas of the sky is obtaining follow-up spectroscopy for every single
source of interest, thereby making sure one has positively identified every member of
the stellar population of interest.
Even with the multiplexing capabilities of modern day multi-fiber
spectrographs, this can be a daunting task, specially if one needs to use very long
exposure times in order to get enough SNR on very faint objects. 
However, in our case, we can extend our study of the 25 Orionis group beyond
the spatial and mass completeness of the spectroscopically confirmed member sample, 
by a judicious use of the photometric candidate member sample. 
At the expense of a larger degree of contamination from non-members, 
the photometric sample has the virtue of providing a spatially complete rendering
of the 25 Orionis aggregate, and it also allows us to sample large numbers of 
objects down to much lower masses.  
The disadvantage of this approach is that
issues like contamination from non-members and variable extinction could
potentially render any reliable analysis as useless.
However, we will show that in the particular case of the 25 Orionis group,
conditions are such that a photometric study of the group properties is not
only feasible, but quite reliable.
%
%
%
%
%
%
\subsection{Interstellar extinction}\label{extintion}
\begin{figure}
\includegraphics[width=80mm]{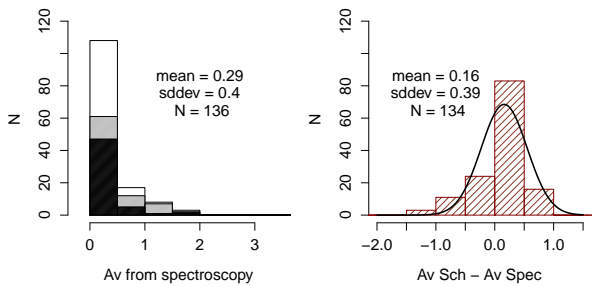}
\caption{Distribution of the visual extinction $A_V$
throughout the 25 Orionis group.
Left panel: $A_V$ values measured from spectra as explained in the
text, from \citet{briceno2005,briceno2007a}
(black), from this work (gray), and the complete
sample (white). Right panel: Distribution of
the residuals between visual extinctions computed
from spectroscopy, and those from the \citet{schlegel1998} 
maps, for the same complete sample of the left panel. Solid line 
indicates the normal distribution
which fits the data, with $\bar{A_V}=0.16$ and $\sigma =0.39$
magnitudes after removing the outliers.}
\label{extinction}
\end{figure}

We determined the visual extinction $A_V$ for each of the 77 spectroscopically 
confirmed members, calculated by using the intrinsic I-J colour for the 
corresponding spectral type proposed by \citet{luhman2003b}, the observed I-J 
colour, and the \citet{cardelli1989} extinction law assuming R$_V=$ 3.09. 
The resulting $A_V$ values are in excellent agreement with previous spectroscopic 
determinations by \citet{briceno2005} for stellar members, as we show 
in Figure \ref{extinction}.

\par
The $A_V$ for the 1246 photometric candidates were derived using the Schlegel 
extinction maps \citep[SEM;][]{schlegel1998}. This approach is valid in this 
particular region because it has a spatially smooth extinction. Still, 
estimates from SEM have two main limitations: First, the $A_V$
reported in the SEM is the integrated extinction along the line
of sight, limited by the sensitivity of the survey and the resulting 
$A_V$ for a nearby object will be overestimated, due to extinction 
sources behind the object. Second, the low spatial resolution of the 
SEM could result in an inadequate interpolation for objects placed 
in regions with abrupt variations of the extinction.
\par
In order to assess how much we would overestimate the extinction
by using the SEM, we computed a mean excess in $A_V$ by
comparing the distribution of the residuals between $A_V$ computed 
using spectroscopy for the members confirmed in this work
and those from \citet{briceno2005},
and $A_V$ from the SEM as we show in Figure \ref{extinction}. 
Except for a few outliers, the resulting distribution is essentially normal, 
with a mean $\bar{A_V}=0.16$ and standard deviation $\sigma_{A_V}=0.39$.
We consider this average $\bar{A_V}$ as the mean excess extinction added by the SEM.
Regarding the issue of limited spatial resolution of the SEM, 
from Figure \ref{figradec1} it is clear that the background towards 
the 25 Orionis group is quite smooth, showing no significant spatial 
variations ($\sigma_{A_V}\sim0.2$).
This uniformity is also confirmed in our Spitzer-MIPS 24$\mu$m images of this area
\citep{hernandez2007} and by preliminary extinction maps performed on the
basis of near IR photometry from VISTA (Alves, private communication).
\par
Thus, we estimated the $A_V$ for each photometric 
candidate by interpolating in the SEM and subtracting 
the $\bar{A_V}$ estimated from the distribution of residuals. 
We consider the mean uncertainty of our estimation over the entire sample 
of candidates as $\sigma_{A_V}/\sqrt{N}=0.03$ magnitudes, where 
$N=134$ is the number of confirmed members used for the
residual distribution.
\subsection{Contamination of the photometric candidate sample by field stars and
extragalactic sources}\label{contamination}
There are three main possible sources of contamination of
a LMS and BD photometric candidate sample from a star forming region:
field dwarf and giant stars, unresolved extragalactic sources and, 
in our case, young stars belonging to 
other sub-regions such as Orion OB1a. Such contamination 
could affect estimates of the IMF, 
the mean age and the spatial distribution of the 25 Orionis group 
when performed using only photometric candidates.
We use a Galactic model to estimate the contamination by Orion non-members 
towards our field, and since we obtained a large number
of spectra, we can verify the predictions from such models.
\par
The number of field stars, of all kinds, in the photometric candidate 
sample, was estimated from a simulation of the expected galactic 
stellar population in an area of $\sim6.4deg^2$ centred at
the 25 Orionis group, performed with the Besan\c{c}on 
Galactic model \citep{robin2003}. 
This model has been extensively tested and correctly reproduces 
the luminosity function within the sensitivity limits of the 2MASS
survey \citep{robin2003}. Second, we reproduce in the Besan\c{c}on 
data the typical photometric incompleteness from our survey according 
to the following procedure: First, we ran the model simulation in the
range 13$<$I$<$22 and simulating the photometric errors as function of 
magnitude, using an exponential function with parameters obtained from 
fitting our survey data. Second, to estimate the fractional 
contamination in our photometric survey we need to remove objects 
from the Besan\c{c}on model in order to reproduce the fall-off as a 
function of magnitude observed in our survey. To do this we 
produced I vs. I-J Hess diagrams for our survey and the simulated sample using 
0.1$\times$0.1 magnitude bins and randomly discarded simulated 
objects in each bin until their number matched our observed number. 
After this correction was applied, a star from the Besan\c{c}on 
simulation was considered a contaminant of the photometric candidate 
sample if it satisfied the selection criteria explained in Section \ref{candidates}. 
\par
\begin{figure*}
\includegraphics[width=80mm]{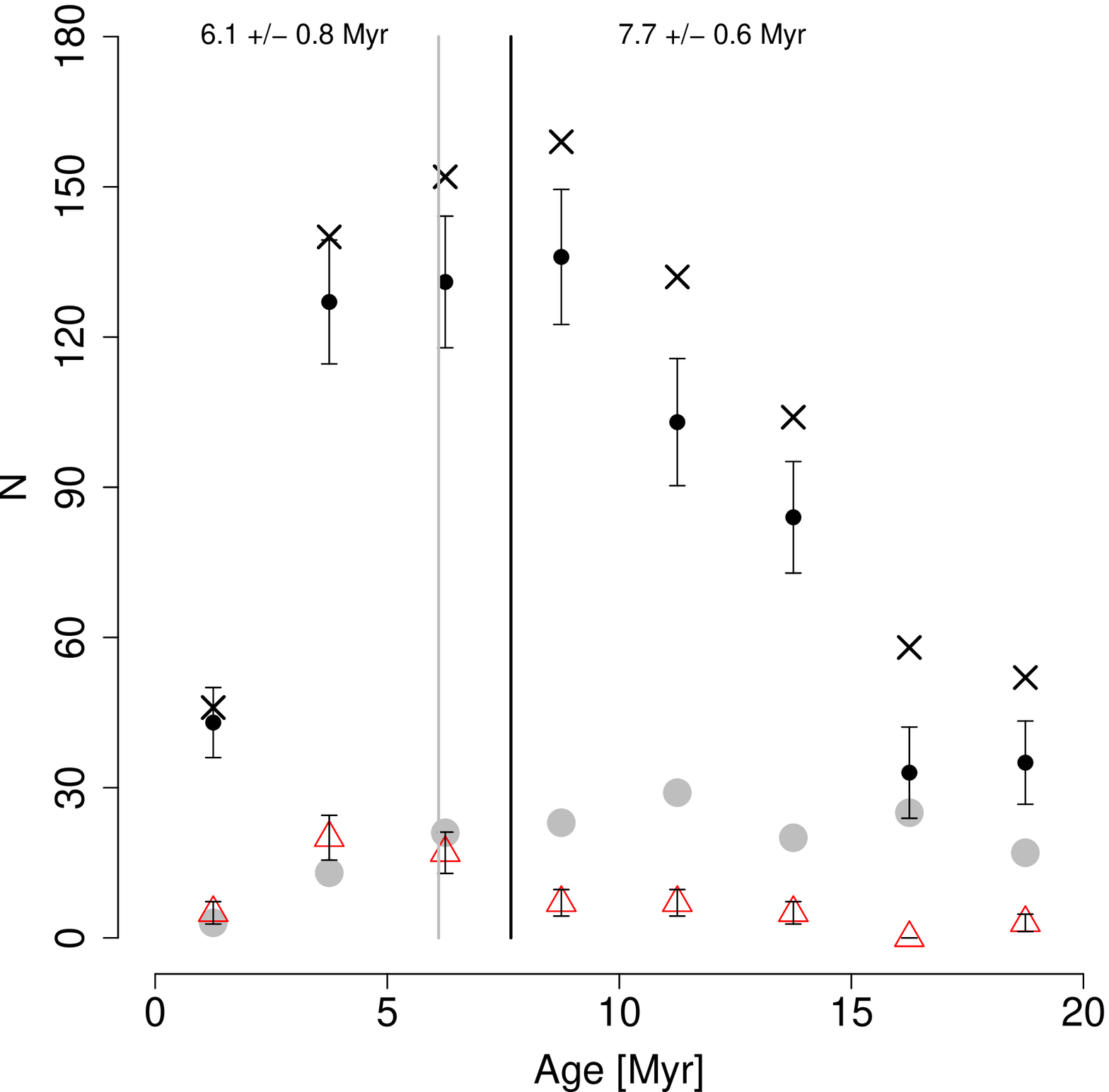}
\includegraphics[width=80mm]{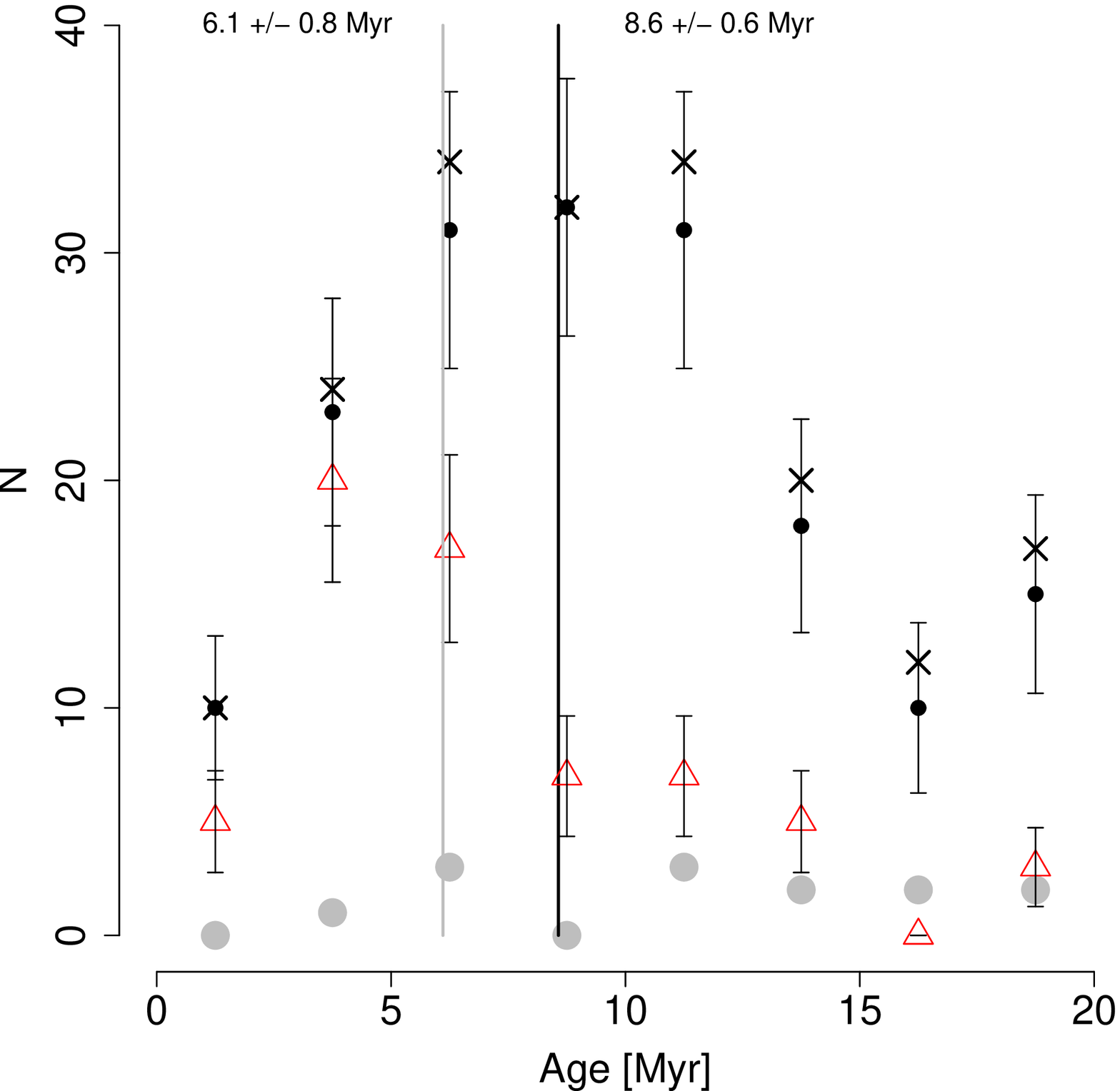}
\caption{Distribution of ages for the full sample of stellar and
sub-stellar candidates and confirmed members in the complete area 
of the survey (left) and inside a circle with 
radius 0.5$^\circ$ centred at the 25 Orionis group 
overdensity defined in Section \ref{spatial} (right).
The distribution corresponding to photometric candidates (crosses) 
was corrected by subtracting the expected contamination by field 
stars (gray circles), resulting in the distribution indicated
with black dots and the corresponding error bars. 
The triangles indicate the distribution for 
the spectroscopically confirmed members from this work and
from \citet{briceno2005}. The distribution for members is the same in 
both plots because the spectroscopic sample covers only the overdensity.
Top labels and vertical lines indicate the mode and standard deviation
of the distributions from confirmed members (left) and from photometric 
candidates (right).}
\label{ages}
\end{figure*}
\begin{figure*}
\includegraphics[width=80mm]{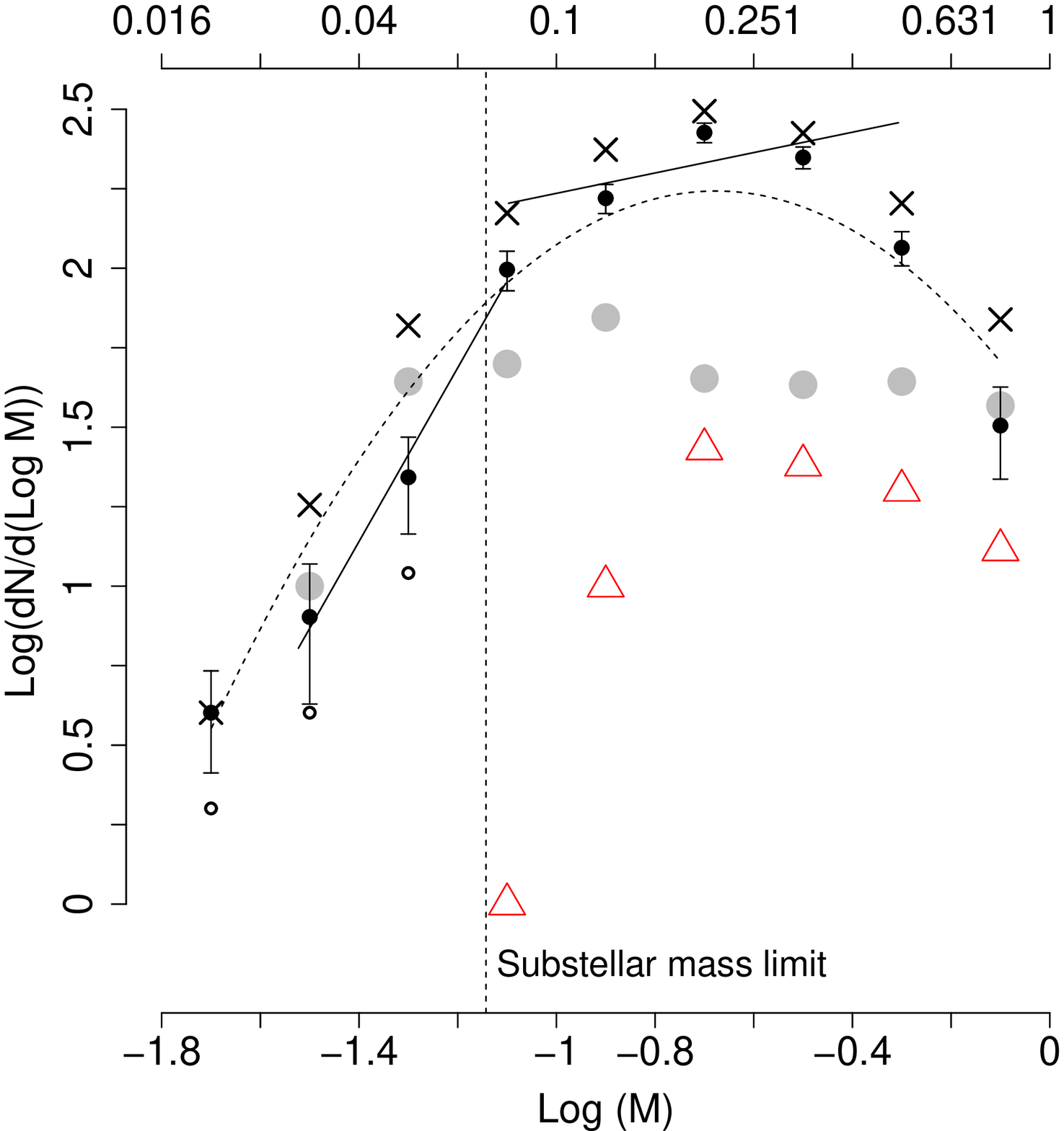}
\includegraphics[width=80mm]{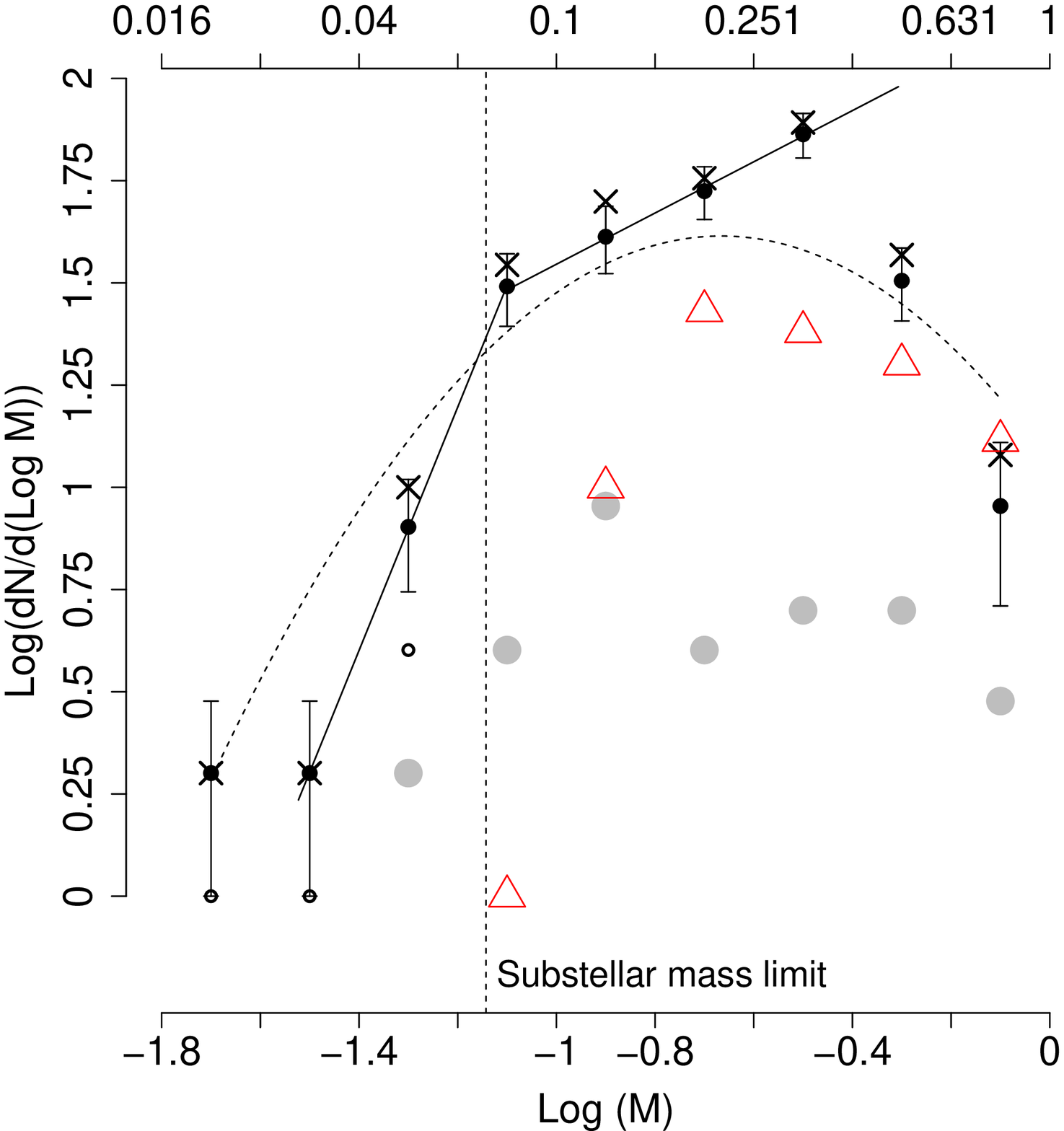}
\caption{The system-IMF for the two samples considered in the 
Figure \ref{ages}. The system-IMF from photometric candidates 
(crosses) was corrected by subtracting the expected contamination 
by field stars (gray circles), resulting in the system-IMF indicated
with black dots and error bars. The empty circles indicate the 
system-IMF before the completeness correction was applied for objects 
with masses $M<0.06M_\odot$. The triangles indicate the system-IMF 
for the spectroscopically confirmed members, which is still largely 
incomplete. We compute all the distributions using bins of 0.2 dex.
The dotted and solid curves indicate the best-fit to a log-normal 
distribution and a Kroupa power-law function.}
\label{imf}
\end{figure*}

The resulting sample of contaminants is constituted exclusively by 212
low-mass main sequence stars that have an homogeneous spatial distribution 
with a mean spatial density of $\sim$33 objects per $deg^2$, estimated 
spectral types from K7 to L0 and distances between $\sim$50 pc and 
$\sim$300 pc. In our spectroscopic sample, every single star rejected as
a member turned out to be a field dwarf, several with some
degree of emission in H$\alpha$ as seen in dMe stars, but all showing
the deep Na I 8183,8195 absorption expected in older, low-mass disc stars.
No late type giants were found among the rejected, non-member, spectroscopic
sample as suggested by our results from Besan\c{c}on.
In order to estimate the contamination as a function of member's mass,
we assume such contaminant field dwarf stars mistaken for members of 
the 25 Orionis group, and we computed the masses and ages they would have,
following the procedure we will explain in Section \ref{agesfim}, resulting 
in the distributions we show in Figures \ref{ages} and \ref{imf}.

We compared the observational J vs J-Ks diagram with those obtained from
an appropiate simulation with the Besan\c{c}on Galactic model
\citep{robin2003} of the expected galactic population in the area covered
by the survey (see Section \ref{contamination}). We found that the point
sources showing J-Ks>1.5 are spatially unresolved extragalactic objects
that are not predicted by the Besan\c{c}on galactic model.
Then, we applyed to the selected extragalactic sources our criteria for the
selection of photometric candidates to members of the 25 Ori group and found
that only a negligible number of unresolved extragalactic sources ($\sim$1
source per $deg^2$) do satisfy our criteria for photometric candidate selection;
this means a total of $\sim$7 extragalactic sources is expected in the
entire sample of photometric candidates.

\par
Summarizing, we found 1246 member candidates in the 25 Orionis group in the mass 
range considered here. If we neglect the contamination due to
extragalactic sources and members from other regions of Orion OB1, from the 
Besan\c{c}on model we expect 316 field dwarfs within the sample for a mean 
density of $\sim$33 contaminants per $deg^2$. The total number of candidates 
expected to be real members of the 25 Orionis group in the mass range considered
here is 930 corresponding to a mean spatial density of $\sim$146 members 
per $deg^2$. Thus, our photometric selection has a global efficiency of 
$\sim$82 per cent which is in good agreement with the 77 new confirmed members 
which corresponds to the $\sim$86 per cent of our spectroscopic survey
because spectra were obtained in the overdensity of candidates
explaines in Section \ref{spatial}. 
We emphasize that the contamination shows a dependence with photometric 
colour which results in the fiducial mass and age distributions shown in 
Figures \ref{ages} and \ref{imf}, which would be the mass and age distributions 
resulting from wrongly considering such contaminants as members. 
%
%
%
%
%
%
\section{Analysis and results}\label{analysis}
\subsection{The mean age and the initial mass function}\label{agesfim}
The masses and ages we derived here for the 25 Orionis low-mass population
were obtained through interpolation of the luminosity and effective temperatures
in the \citet{baraffe1998} models, adopting a distance of 360 pc from 
Brice\~no et al. (2014, in preparation).
For the photometric candidates we interpolated the luminosity 
and the effective temperature we obtained by interpolation of the 
I-band magnitude and the I-J photometric colour, both corrected 
by interstellar extinction as explained in Section \ref{extintion}, 
using the relationships from \citet{luhman2003b} and \citet{dahn2002},
from which we obtained the effective temperatures 
and bolometric corrections. For the spectroscopically confirmed 
members we followed two procedures: First, as we did for the 
photometric candidates. Second, by interpolation of the measured 
spectral types in the same relationships. Within the uncertainties, 
no significant differences were found for masses and ages derived 
with either procedure for the spectroscopically confirmed member sample.
This similarity is not surprising, given the low and smooth interstellar 
reddening towards this region, and the $\sim$0.5 spectral subclass mean 
uncertainty in the estimation of spectral types.
\par
We derived the distribution of ages and the system-IMF for the photometric
candidates and for the confirmed members in the entire area of our
survey and in the spatial overdensity that will be defined
in Section \ref{spatial} as the 25 Orionis group overdensity.
Figure \ref{ages} shows the resulting distribution of ages for 
photometric candidates and for spectroscopically confirmed members
including also the members from \citet{briceno2005} within the
same mass interval. The member sample is still largely incomplete 
and biased toward the brighter (more massive) objects, but as can be seen in 
Figure \ref{ages}, this bias does not strongly affect
the estimation of age. 
The distribution corresponding to photometric
candidates was corrected by subtracting the age distribution
for the expected contamination by field stars, i.e.
the age distribution that results from mistaking the contaminants
as members of the 25 Orionis group. 
\par
We found a good agreement between the distributions of ages
computed from photometric candidates and from spectroscopically
confirmed members.
In the complete area of the survey, the derived age of the 25 Orionis 
group is $7.7\pm0.6$ Myr from candidates and 
$6.1\pm0.8$ Myr from confirmed members. In the spatial overdensity,
where the new members were detected, we found a mean age of
$8.6\pm 0.6$ Myr from candidates. All the ages reported correspond to
the mode of each distribution. We adopt the age derived from the 
confirmed members ($6.1\pm 0.8$ Myr) as the age of the 25 Orionis group, 
although this value is statistically indistinguishable from that obtained 
from photometric candidates.
Therefore we confirm the age derived in previous works by our group 
\citep[$\sim$7 Myr ;][]{briceno2007a}. In this new sample 
we are including several new spectroscopically confirmed 
members and photometric candidates covering a much wider 
magnitude range which allows for a more robust estimation 
of the mean age. Additionally, our result shows that the 
population within the 25 Orionis group is not distinguishable 
from the surrounding population in terms of their ages alone. 
\par
Figure \ref{imf} shows the system-IMF in the LMS and BD regime for the photometric 
candidates of the 25 Orionis group. 
Although the photometric completeness of our survey (I=19.6) is equivalent
to a completeness mass limit of 0.03 $M_\odot$ assuming the \citet{baraffe1998}
models with an age of 7 Myr old and a distance modulus of m-M=7.78, the 
width of the membership loci defined in Section \ref{candidates}
and the dispersion of the candidates within it, suggest that 
some objects more massive than 0.03 $M_\odot$ could be located below the 
photometric completeness. In order to account for such incompleteness
in our sample we estimated, per magnitude bin, the number fraction of 
missing objects as the ratio between the expected and observed number of objects 
fainter than the completeness magnitude\footnote{The observed  number counts 
(for the full catalogue) behave linearly on a log scale up to
the completeness magnitude (by definition), after which these start
to decline. We extrapolate this linear behaviour for magnitudes fainter
than the completeness, and use this fit to compute the expected number counts.}.
Figure \ref{imf} shows the system-IMF obtained after this
completeness correction and the correction for the contamination 
by field stars were applied.
\par
Because we did not apply any correction to account for unresolved
binaries, what we report is the so called \emph{system-IMF} of 
the 25 Orionis group \citep[see e.g.][and references therein]{parravano2010}. 
Although the system-IMF does not reveal by itself the real number of
objects per mass bin, it is useful for comparison with 
system-IMFs from other SFR and young clusters, assuming that 
the samples have similar spatial resolution,
and that the populations have similar multiplicity properties
such as the number fraction of objects belonging to multiple systems, 
separation and mass ratio distributions.
\par
\begin{figure}
\includegraphics[width=80mm]{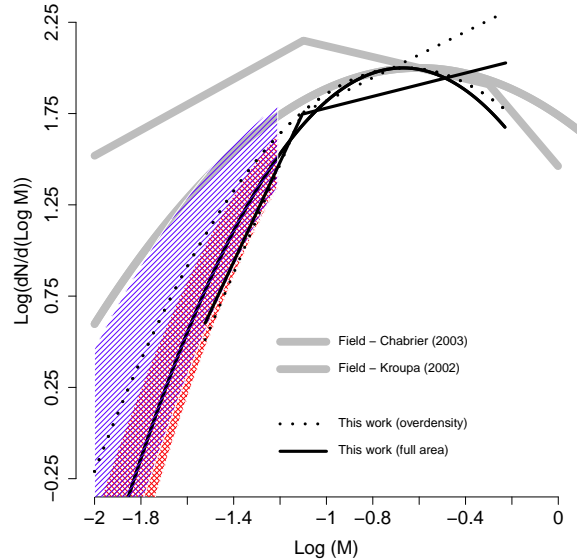}
\caption{Comparison of the system-IMF obtained for the 25 Orionis 
group with those from \citet{chabrier2003b} and \citet{kroupa2001,kroupa2002}. 
The black solid curve and straight lines correspond to the fit to a log-normal and 
power-law functions for the entire area of the survey respectively. The gridded
area indicates, in the BD regime, the uncertainty in the fit to the log-normal 
function.
The dotted curve and straight lines: fit to a log-normal and
power-law functions for the spatial overdensity. The dashed
area indicates, in the BD regime, the uncertainty in the fit to the log-normal 
function.
Gray curve: IMF from \citet{chabrier2003b}. Gray solid lines: IMF 
from \citet{kroupa2001,kroupa2002}.}
\label{imfcomp}
\end{figure}
\begin{figure}
\includegraphics[width=80mm]{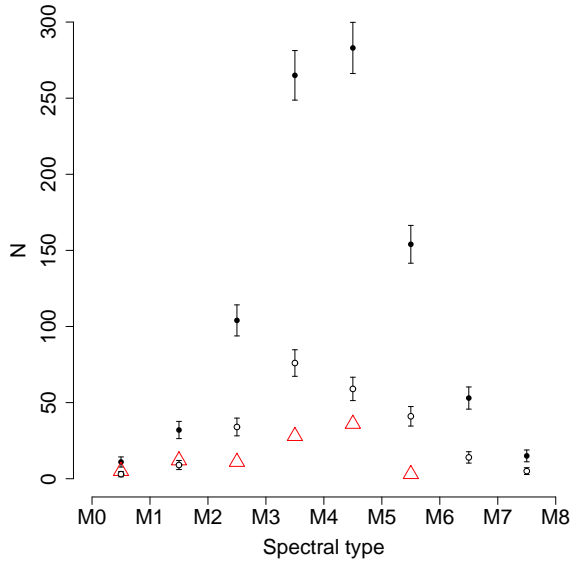}
\caption{The distribution of spectral types in the 25 Orionis group. 
Triangles: confirmed members from this work and from \citet{briceno2007a}. 
Black and empty dots indicate respectively the candidates
in the entire area of the survey and inside the overdensity
previously corrected by the expected contamination by field stars.}
\label{stdis}
\end{figure}

We describe the derived system-IMF according to a Kroupa power-law 
function $\Psi(m) = dN/dm \propto m^{-\alpha_i}$ \citep{kroupa2001,kroupa2002},
and a Scalo log-normal function \citep{scalo1986} in the form
$\Psi(\log m) = dN/d(\log m) \propto exp{-((\log m-\log m_c)^2/2\sigma^2)}$
\citep{chabrier2005}. We fitted the mass distributions within the mass range 
$0.03<M/M_\odot<0.8$ to the log-normal function and within the mass ranges
$0.03<M/M_\odot<0.08$ and $0.08<M/M_\odot<0.5$ for the power-law description.
As we did for the age estimate, we obtain the system-IMF for the candidates
inside the spatial overdensity (see Section \ref{spatial})
and for the entire area of the survey.
For the latter, the system-IMF for masses in the intervals $0.03<M/M_\odot<0.08$ and
$0.08<M/M_\odot<0.5$ follows Kroupa-type functions with $\alpha_3 = -1.73\pm0.31$ 
and $\alpha_2 = 0.68\pm0.41$ respectively, and that in the complete mass interval
$0.03<M/M_\odot<0.8$ is well described by a log-normal function
with coefficients $m_c = 0.21^{+0.02}_{-0.02}$ and $\sigma = 0.36\pm0.02$. 
For the spatial overdensity we found $\alpha_3 = -1.97\pm0.02$, $\alpha_2 = 0.37\pm0.04$,
$m_c = 0.22^{+0.02}_{-0.02}$ and $\sigma = 0.42\pm0.05$.
In Figure \ref{imf} and Table \ref{table:imf} we show the resulting 
fits and in Figure \ref{imfcomp} we compare our derived IMF 
with other IMFs from the literature in which no correction for
binarity was made.
\begin{table*}
\centering
\begin{minipage}{160mm}
\caption{Fit parameters for the system-IMF.}
\begin{tabular}{lcccccc}
\hline
Population & M interval & $\alpha_2$ & $\alpha_3$ & $m_c$      & $\sigma$   & Reference \\
\hline
25 Orionis overdensity & $0.03<M/M_\odot<0.8 $ & $0.37\pm0.04$  & $-1.97\pm0.02$ & $0.22^{+0.02}_{-0.02}$      & $0.42\pm0.05$ & This work \\
25 Orionis entire area & $0.03<M/M_\odot<0.8 $ & $0.68\pm0.41$  & $-1.73\pm0.31$ & $0.21^{+0.02}_{-0.02}$      & $0.36\pm0.02$ & This work \\
Field                  &      $M<M_\odot$      &      \dots     &  \dots         & $0.22$                      & $0.57$        & \cite{chabrier2003b} \\
\hline
\end{tabular}
\label{table:imf}
\end{minipage}
\end{table*}
Additionally, we computed the distribution of spectral types, which is 
generally used as a proxy of the IMF which avoids most of the 
uncertainties related to the computation of masses
\cite[e.g.][]{preibisch2012}. Figure \ref{stdis} shows the 
spectral type distributions for confirmed members and photometric 
candidates. For the later, the spectral types were estimated from 
the observed $I-J$ colour, dereddened following the procedure explained 
in Section \ref{extintion}, and applying the colour-spectral-type 
relationship from \citet{luhman2003b}. 
The distribution of spectral types for candidates
peaks around $\sim M4.5$ and declines for earlier and
latter objects, and is consistent with the spectral type distributions reported
for confirmed members in other regions such as IC348 and
Chamaeleon I \cite[e.g][]{luhman2012}.
\par
The system-IMF, and the spectral type distribution are commonly 
used as observables to constrain which processes and conditions 
are preponderant during the formation of 
LMS and BDs. In the sub-stellar domain some variations have been 
observed in these quantities, Taurus probably being the most
representative case \citep{briceno2002}. 
These variations are not well understood and more observations 
are needed in order to better constraint these phenomenon. What we have presented 
here is a new photometry-based system-IMF in the mass range
$0.03<M/M_\odot<0.8$, obtained for a $\sim7$ Myr old dispersed population 
in a low-density environment, using a numerous sample with 1246 objects.
The peak of the spectral-types distribution and the characteristic mass 
$m_c$ of the log-normal representation of the system-IMF we found for 
the 25 Orionis group, is consistent with those reported for other SFR 
and young clusters \citep[e.g.][]{oliveira2009}, which, as mentioned 
by \citet{oliveira2009}, supports the idea of an IMF relatively 
independent of environmental properties such as density, temperature, 
metallicity and radiation field, as predicted by the models from 
\citet{bonnell2006} and \citet{elmegreen2008}. 
On the other hand, the standard deviation $\sigma$ from the log-normal 
fit results to be slightly smaller than is expected from \citet{chabrier2003b} 
which is also consistent with the small values for the $\alpha_3$ coefficients 
obtained from the fit to a power-law IMF. This suggest that the 25 Orionis 
population could have a lower number of BDs. 
Because the fits are strongly dependent on small variations in the number of 
objects in the least massive bins, more observations are needed in order
to better constraint the system-IMF in the sub-stellar domain of the 25 
Orionis group and confirm such a trend.
A sensitive photometric survey for the detection of a complete
sample of BDs down to $\sim0.01 M_\odot$, and the spectroscopic 
confirmation of members in the entire mass range (including also 
$M>0.8M_\odot$) of this group are needed to confirm this low number of BD. 
Also, kinematic information is needed in order to evaluate whether the 
dynamical evolution of the group could be biasing the result. High spatial
resolution imaging intended to detect binary or multiple systems
are important in order to establish the effect that such
a population of binaries could have on the observed system-IMF.
Spectroscopic observations of our least massive candidates and
much deeper photometric observations are
currently under way.
\subsection{The spatial distributions}\label{spatial}
\begin{figure}
\includegraphics[width=80mm]{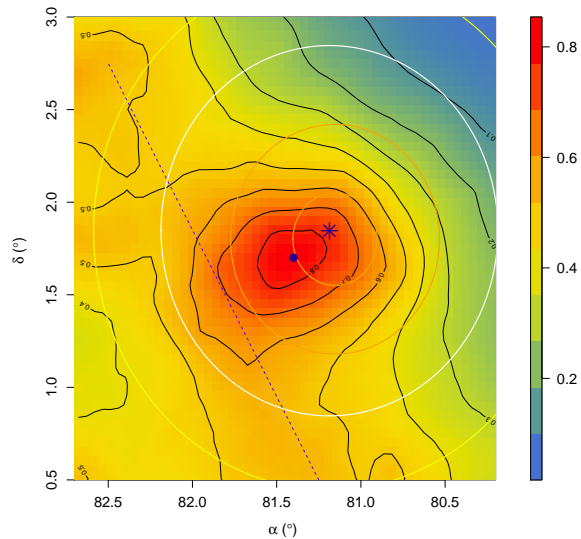}
\caption{Spatial density distribution of the LMS and BD photometric
candidates of the 25 Orionis group and its surroundings.
Colored scales and isocontours indicate the mean
density computed in bins of $3^\prime \times 3^\prime$ and smoothed
with a gaussian with standard deviation of $6^\prime$ in both directions.
The central strong overdensity is the 25 Orionis group and the doted line
indicates the filamentary overdensity explained in the text.
The circles indicate the radii from the previous works shown also in
Figure \ref{figradec1}. The contamination from field stars present
in the candidate sample is spatially smooth.
The central big dot indicates the centre of the group computed in this 
work, and the big asterisk, the position of the star 25 Ori.}
\label{figspatial1}
\end{figure}

Here we present the first comparative study of the spatial distribution
of $\sim$1100 candidate LMS and $\sim$120 candidate BDs within the area
covered by the 25 Orionis group and its surroundings.
With this deep and spatially complete photometric sample we
confirm the 25 Orionis group as the strong overdensity of stellar
\emph{and} sub-stellar candidate and members shown in Figure \ref{figspatial1}.
We compute the number of objects on $3^\prime \times 3^\prime$ bins and
smoothed the resulting pattern with a gaussian function with standard deviation
of $6^\prime$ in both spatial directions.
The overdensity is centred at $(\alpha,\delta)\sim(81\fdg2,1\fdg7)$
with a mean radius of $\sim0\fdg5$ and a slight elongation approximately
in the East-West direction. Its mean density is of $\sim$0.5 objects per
$3^\prime \times 3^\prime$ bin ($\sim200 obj/deg^2$), surrounded by a typical
mean density of $\sim$0.25 objects per $3^\prime \times 3^\prime$ square
($\sim100 obj/deg^2$).
The centre was computed as the position of the maximum density within the
overdensity and the mean radius as the distance from the centre at which
the density drops to the value of the mean density of the surroundings.
Assuming a distance of $\sim$360 pc the angular radius corresponds to a
mean linear radius of $\sim$3.14 pc.
We note that the 25 Orionis overdensity appears to be superimposed on a
low-density diagonal filament indicated in Figure \ref{figspatial1},
whose mean density is $\sim$0.3 objects per $3^\prime \times 3^\prime$ square.
\par
The overdensities could be intrinsic to the population, a
consequence of observational biases or a combination of both.
We consider the detected overdensities as real because:
(i) They are not a consequence of interstellar extinction, since
the latter does not significantly affect the number of selected
candidates and the SEM does not shows a spatial distribution
similar with those from candidates, as can be seen in Figure \ref{figradec2}.
(ii) They cannot be explained as a consequence of contamination
by field stars because these follow a smooth spatial distribution
which depends on galactic latitude as predicted by
the Besan\c{c}on model.
(iii) They are not a consequence of a photometric
bias because the completeness and limiting magnitudes
in all bands used for the candidate selection
are spatially homogeneous in the area where the
overdensities were detected and the slight difference in the completeness magnitude
within observations centred at $\delta=1^\circ$
and $\delta=3^\circ$ (Table \ref{coaddedscans})
cannot produce the detected overdensities.
(iv) They are not a consequence of any artifact in the
images or inappropriate identification
of sources from different catalogues as
we concluded from visual inspection of the images.
\par
\begin{figure*}
\includegraphics[width=80mm]{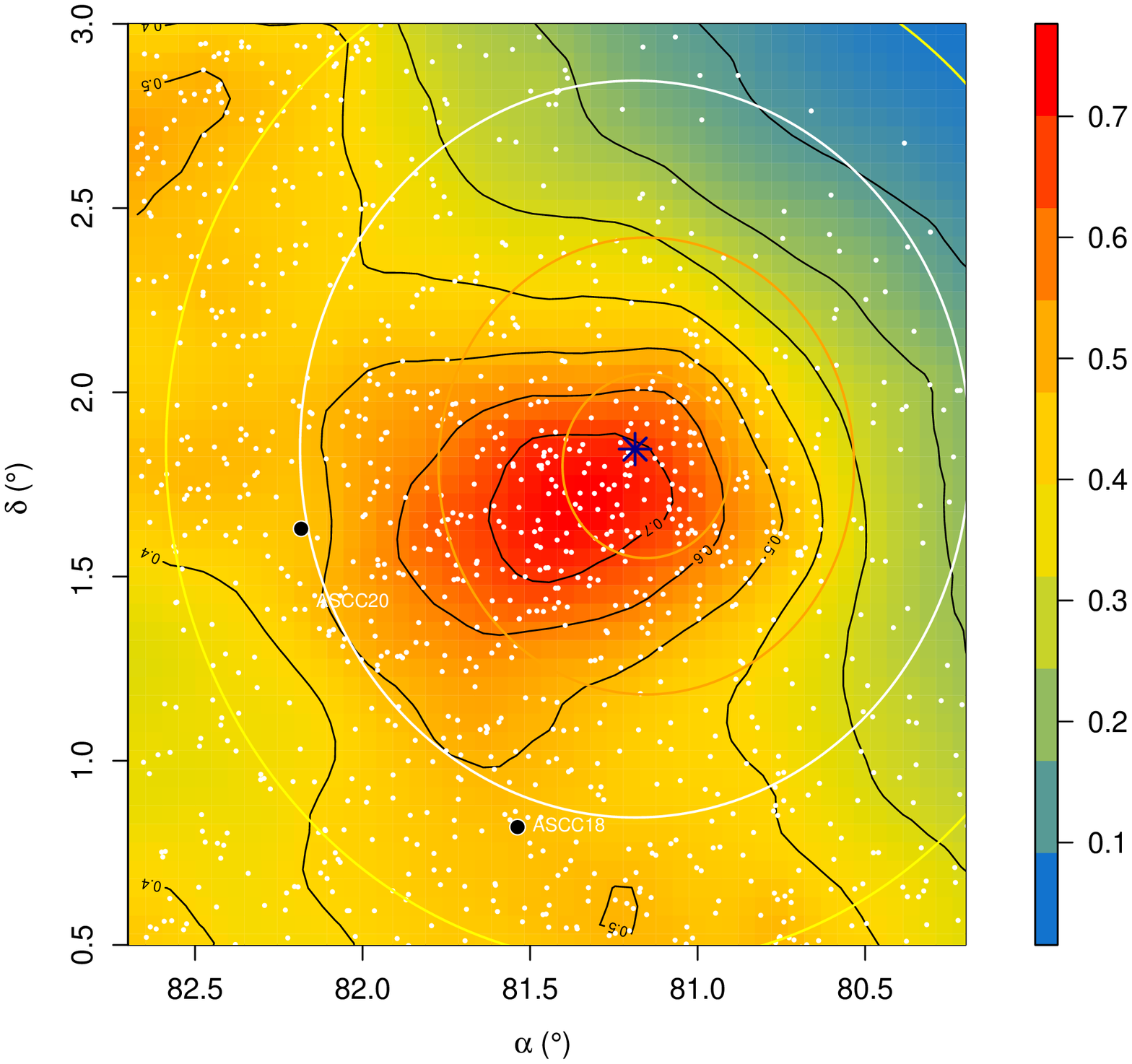}
\includegraphics[width=80mm]{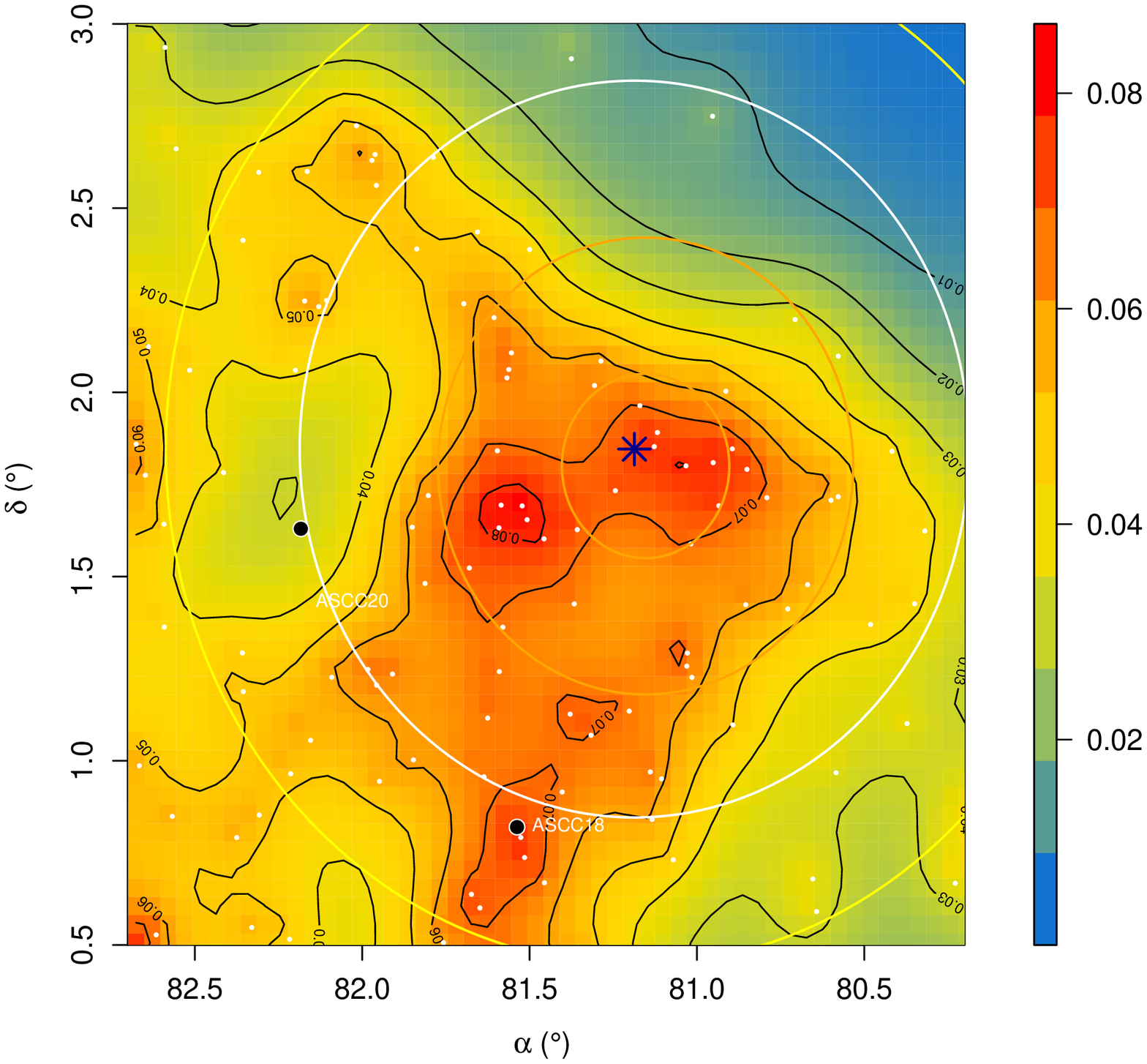}
\caption{Comparison of the density distributions of LMS candidates
($M>0.08M_\odot$, white dots, left) and of BD candidates
($M<0.08M_\odot$, white dots, right). The K-S test probability
between both distributions is 0.30 which indicates that both
distributions are statistically indistinguishable.
The big central asterisk represents the 25 Ori star and
the black dots clusters from \citet{kharchenko2005}.
Other elements are as in Figure \ref{figspatial1}.}
\label{figspatial2}
\end{figure*}

The East-West elongation of the 25 Orionis overdensity is suggested from the spatial
distribution of the members confirmed by \citet{briceno2005} and also by the
spatial distribution of the high mass candidates of \citet{kharchenko2005} with
the highest probability of membership.
In this work we confirm this trend with
a numerous sample of well characterised candidates covering
not only the 25 Orionis group but a reasonable area of its surroundings.
Furthermore, our results rule out the idea of a possible extension of the
25 Orionis group to the South as suggested by \citet{mcgehee2006}.
With this new data set, deeper and covering the entire
area surrounding the group, we find that the
possible southern extension of the group reported by \citet{mcgehee2006}
is probably part of the filament described above that extends
also to the North-East direction, or the cluster ASCC18 from \citet{kharchenko2005}. 
This conclusion could not have been
derived from the data considered by \citet{mcgehee2006}, which
covers only the southern part of the area shown in Figure \ref{figradec1}.
We note that the southern overdensity detected by \citet{mcgehee2006}
at $(\alpha,\delta) \sim (81\fdg2,0\fdg8)$ is also observed in
Figures \ref{figspatial1} and \ref{figspatial2} as part of the 
southern extension of the filament (note the different scales 
between our plots and Figure 8 from \cite{mcgehee2006}), but 
its mean low-density and position do not suggest it is part of 
the overdensity we consider here as the 25 Orionis group.
\par
Results from \citet{van2006} show that during their first
$\sim$15 Myr, $\sim$20 per cent of galactic clusters and associations
have diameters larger than 7 pc, being expanding associations
rather than gravitationally bound clusters.
Therefore the diameter obtained in this work for the 25 Orionis
group (D$\sim$6.3 pc) suggests it could be a bound cluster given
its age of about $\sim$7 Myr. Nevertheless, this computation
is based on the radius found for the spatial overdensity,
but we cannot rule out the cluster is spread over a larger area,
since, as noted in Section \ref{agesfim}, the age distribution
and system-IMF are indistinguishable from its surroundings.
A formal computation of the virial mass of this cluster is
out of the scope of this work, since it depends almost entirely
on its massive stars, which are not discussed here.
\par
The panels of Figure \ref{figspatial2} show separately the spatial
density distribution of $\sim$1100 LMS and $\sim$120 BDs.
In order to test whether the observed spatial distributions for BDs 
and LMS are drawn from the same parent distribution (null hypothesis), 
we have computed a two-dimensional two-sample Kolmogorov-Smirnov (K-S) 
test \citep{press2007}. For the full BD and LMS samples the K-S
statistic results in a probability $P=0.30$ of the null hypothesis 
being true, which means the difference between both spatial distributions 
is not statistically significant \citep{press2007}. In order to assess 
the robustness of this result, we generated $10^3$ bootstrap resamples 
of both the BD and LMS distributions.
A single bootstrap sample is made, e.g. for BDs, by randomly sampling
with replacement from the BD catalogue,
until a new set with the same number of objects (as the original
catalogue) is generated. This is a standard
technique that allows producing a set of random realizations drawn
from the same distribution as the parent
populations, which in this case are the BD and VLM samples \citep[see
e.g.][for more details]{wall2012,press2007}.
We computed the same K-S test on the $10^3$ bootstrap BD and LMS
samples and found
that only in 12 per cent of the experiments the distributions turned out
to be incompatible at the 99 per cent confidence level ($P<0.01$).
This shows it is a robust result, independent on stochastic
fluctuations due to sample size, that the spatial distributions
observed for BDs and LMS in this 7 Myr old population are
statistically indistinguishable.
\par
Although this result is not consistent with the mass segregation 
predicted by the early premature ejection models from \cite{reipurth2001} 
subsequent work by \cite{bate2003} does not predict significantly different 
velocity dispersions for ejected BDs and stars. Thus, the similar
spatial distributions do not necessarily reject the premature ejection
scenarios. In any case, what we are presenting here constitutes a new 
observational evidence supporting the similarity of the spatial 
distribution of LMS and BDs for a $\sim$7 Myr old population.
\subsection{Disc indicators and accretion signatures}\label{discs}
\begin{figure*}
\includegraphics[width=80mm]{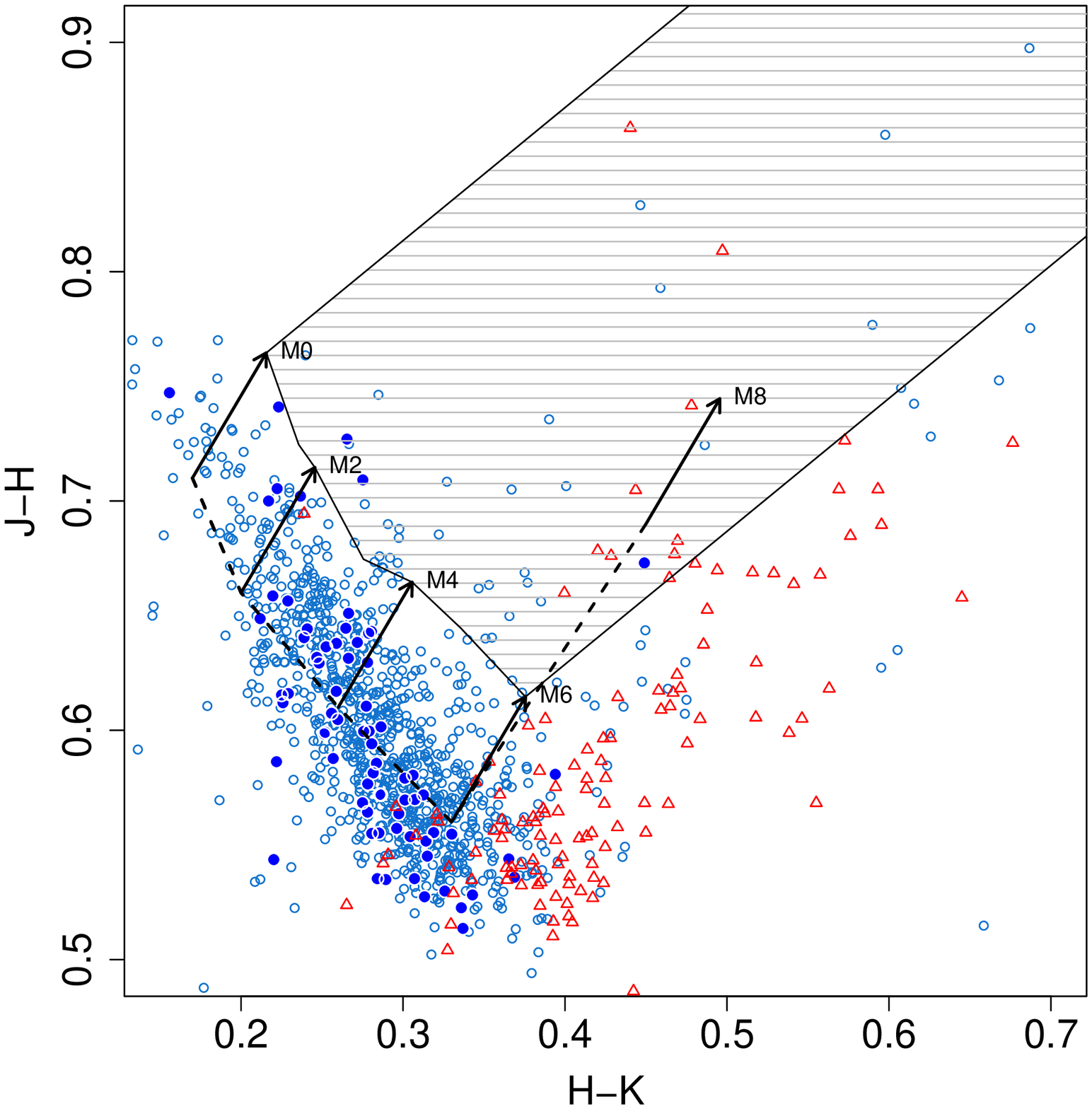}
\includegraphics[width=80mm]{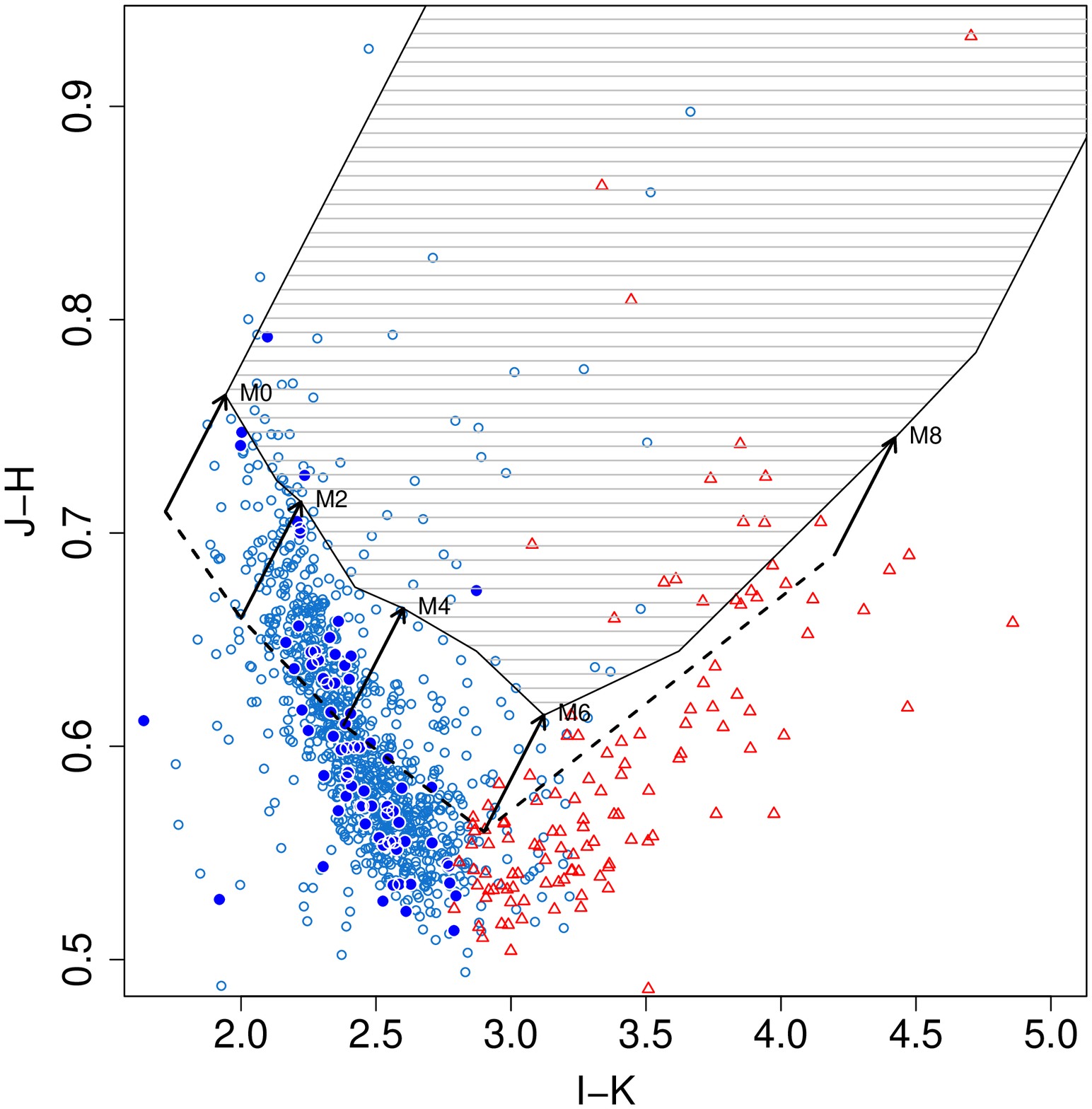}
\caption{Optical and near-infrared colour-colour diagrams of
spectroscopically confirmed members and photometric candidates.
Left panel: J-H vs. H-Ks. Right panel: I-Ks vs. J-H.
The dashed lines indicates the photospheric loci from
\citet{luhman2003b} and the shaded areas indicate the excess
loci explained in Section \ref{discs}. Empty and filled circles
indicate respectively candidates and confirmed LMS.
Triangles indicates candidate BDs. Reddening vectors for
several spectral types are also indicated for the typical mean
visual extinction ($A_V\sim0.5$ mag) through the 25 Orionis
group. Number fractions of objects showing IR-excesses
are listed in Table \ref{table:discfrac}.}
\label{ircc1}
\end{figure*}

We computed the number fraction of LMS and BDs showing near
IR excesses as an indication of the presence of primordial 
circumstellar inner-discs. In the left panel of 
Figure \ref{ircc1} we show a J-H vs. H-Ks colour-colour diagram 
with the CTTS locus from \citet{meyer1997} and its extrapolation
to M6 spectral type.
We obtained that $3.8\pm0.4$ per cent of LMS candidates
and $4.5\pm0.5$ per cent of spectroscopically confirmed LMS
are placed inside the CTTS locus.
Both results are in good agreement within the uncertainties
and are consistent with those from 
\citet{briceno2007a} ($\sim6$ per cent). The fractions reported here
for the photometric candidates were computed considering
the expected contamination by field stars and the errors
were computed according to the binomial distribution.
\begin{table*}
\centering
\begin{minipage}{110mm}
\caption{Number fraction of LMS and BDs showing infrared excesses.}
\begin{tabular}{cccc}
\hline
Diagram & LMS members  & LMS candidates & BD candidates \\
        & [per cent]   & [per cent]     & [per cent]    \\
\hline
J-H vs. H-Ks                                     & $4.5\pm0.5$  & $3.8\pm0.4$ & $\cdots$      \\
I-Ks vs. J-H                                     & $4.1\pm0.4$  & $5.6\pm0.5$ & $12.7\pm0.5$  \\
3.6$\mu m$-4.5$\mu m$ vs. 4.5$\mu m$-5.8$\mu m$  & $7.5\pm0.4$  & $7.8\pm0.3$ & $18.1\pm0.3$   \\
3.6$\mu m$-4.5$\mu m$ vs. 5.8$\mu m$-8.0$\mu m$  & $12.0\pm0.4$ & $9.4\pm0.8$ & $39.2\pm0.4$  \\
Ks-8.0$\mu m$ vs. Ks-5.8$\mu m$                  & $5.7\pm0.4$  & $7.9\pm0.5$ & $47.0\pm0.5$  \\
\hline
\end{tabular}
\label{table:discfrac}
\end{minipage}
\end{table*}
\par
For objects later than M6, the J-H vs. H-Ks
diagram is not a good choice for detecting
infrared excesses because the photospheric locus
between spectral types M6 and M9 is essentially colinear with
the corresponding reddenning vectors, which prevents a reliable 
measurement of the fraction of objects showing IR excesses as a 
function of their spectral-types.
Because of this limitation, we apply an empirical criterion to detect 
objects with infrared excesses, using the I-Ks vs J-H diagram,
where the photospheric loci and reddening vectors for spectral
types between M6 and M9 are more separated, as we show in the 
right panel of Figure \ref{ircc1}.
Essentially we define a locus whose lower limit in the I-Ks vs J-H 
diagram is parallel to the photospheric locus from \citet{luhman2003b}, 
and which is separated from it by a distance equivalent to a 
reddening vector corresponding to $A_V=0.5$, which is slightly
larger than the mean value of the interstellar extinction across 
the 25 Orionis group, as we have shown in Section \ref{extintion}. 
We found that $\sim 5.6 \pm 0.5$ per cent of the 
candidate LMS and $\sim 4.1 \pm 0.4$ per cent of the confirmed LMS show
excesses. 
For the substellar regime the number fraction of candidates
showing such IR excess is $12.7\pm0.5$ per cent.
Again, the number fraction of stars showing infrared excesses
from inner discs agrees well with previous determinations by
\citet{hernandez2006} and \citet{briceno2007a} using Spitzer data. 
In the sub-stellar regime, our result is the first estimate 
obtained for the 25 Orionis group.
\par
\begin{figure*}
\includegraphics[width=80mm]{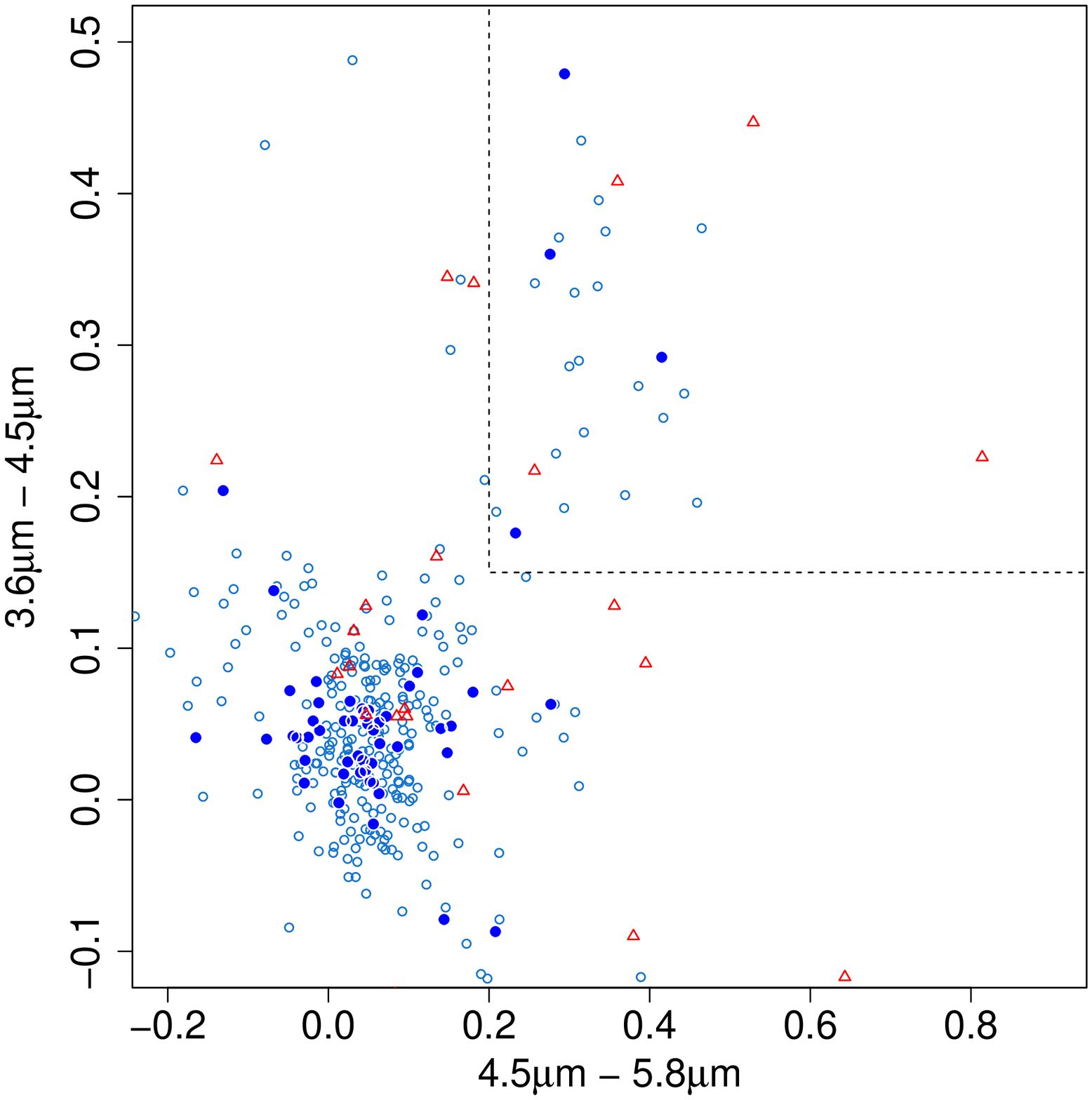}
\includegraphics[width=80mm]{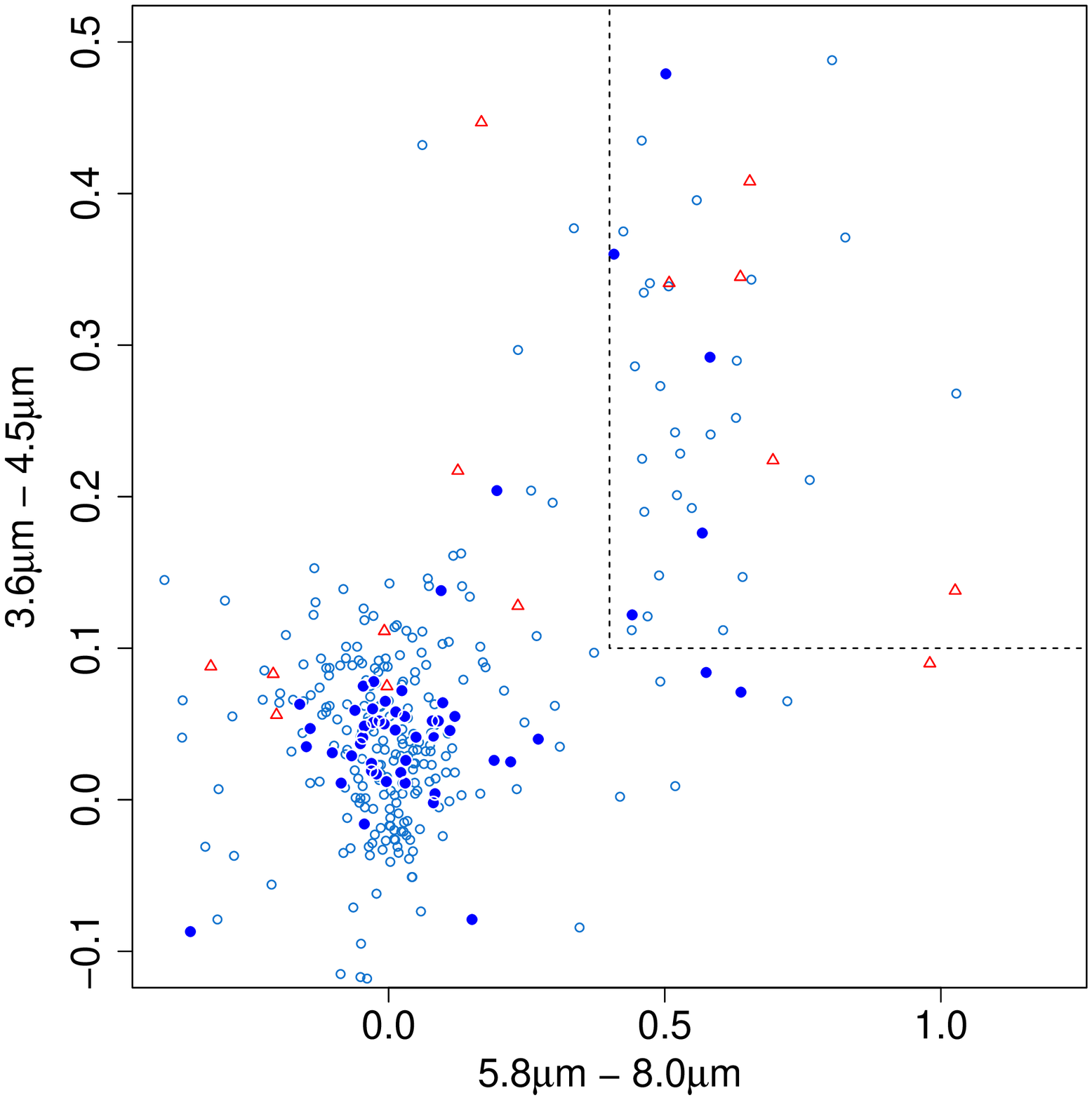}
\caption{IRAC colour-colour diagrams of spectroscopically confirmed members
and photometric candidates. Symbols are as in Figure \ref{ircc1}.
Left panel: the 3.6$\mu$m-4.5$\mu$m vs. 5.8$\mu$m-8.0$\mu$m diagram showing the
CTTS locus (dashed lines) defined by \citet{hartmann2005b}.
Right panel: the 3.6$\mu$m-4.5$\mu$m vs. 4.5$\mu$m-5.8$\mu$m diagrams with the excess
region (dashed lines) defined by \citet{luhman2005}.
Number fractions of objects showing IR-excesses
are listed in Table \ref{table:discfrac}}
\label{memberscc2}
\end{figure*}
\begin{figure*}
\includegraphics[width=80mm]{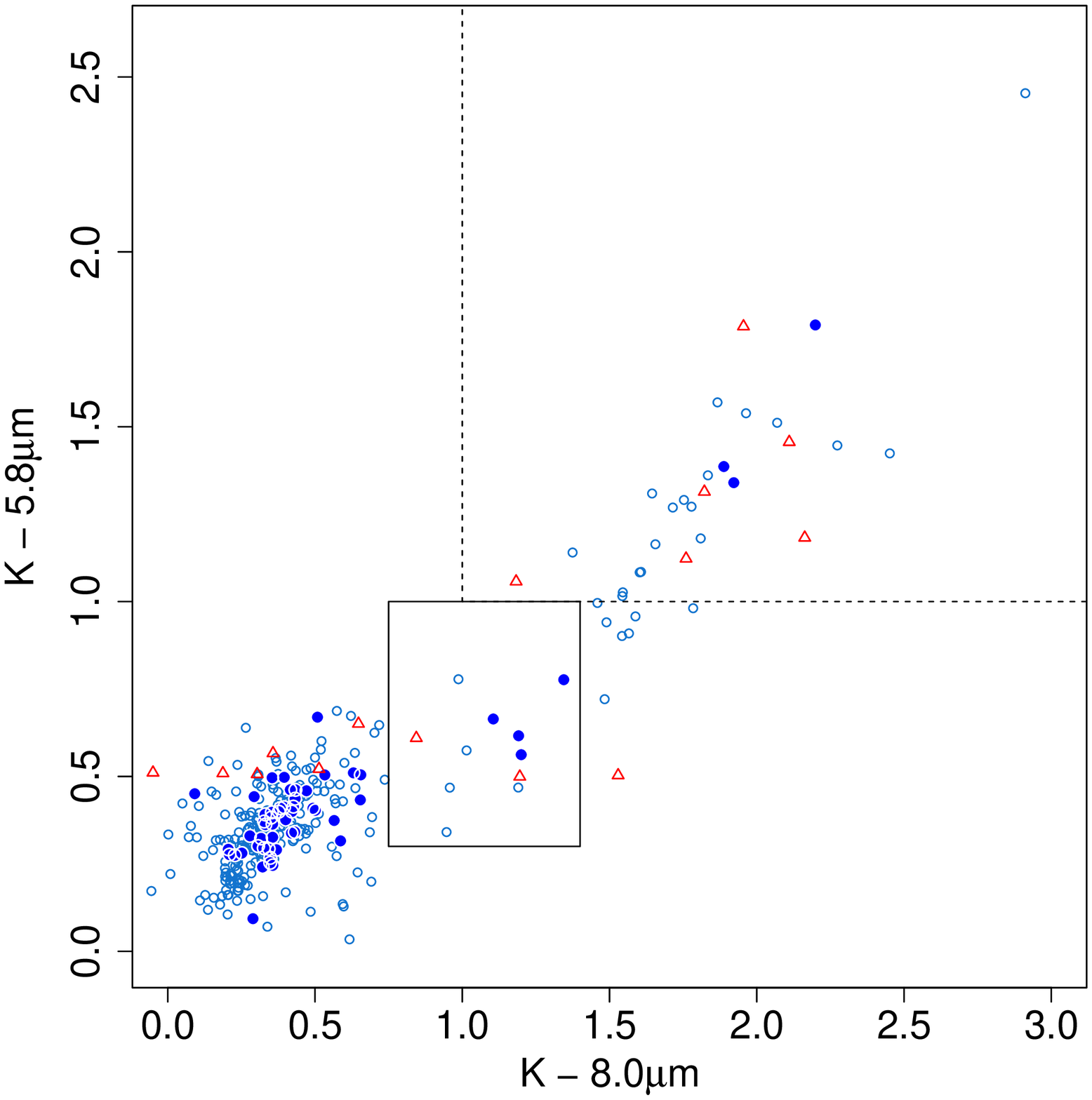}
\includegraphics[width=80mm]{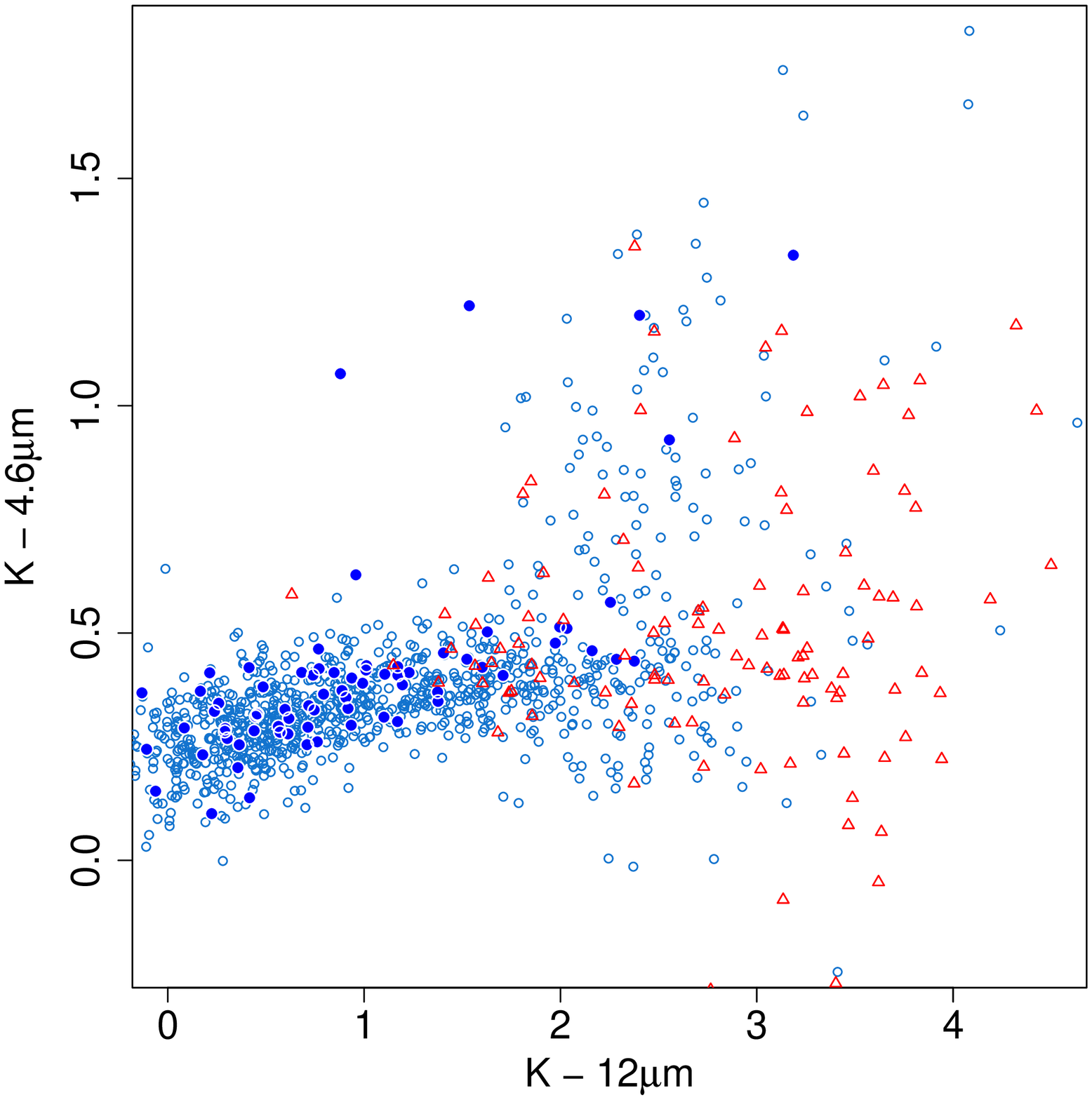}
\caption{Color-colour diagram combining photometry from VISTA,
IRAC and WISE. Symbols are as in Figure \ref{ircc1}.
Dashed polygons indicate disc loci and the solid polygon
in the left panel encloses the possible transitional or
evolved discs from the colour intervals defined by
\citet{luhman2010}.}
\label{memberscc3}
\end{figure*}

Emissions from warm dust in evolved discs or in the external regions of 
primordial discs, occurs at longer wavelengths. In Figures 
\ref{memberscc2} and \ref{memberscc3} we show colour-colour
diagrams in the wavelength range from $\sim2.2\mu$m to 
$\sim12\mu$m, for confirmed members and photometric candidates 
with magnitudes from VISTA, IRAC and WISE.
The left panel of Figure \ref{memberscc2} shows the 3.6$\mu$m-4.5$\mu$m vs. 4.5$\mu$m-5.8$\mu$m
diagram from IRAC data with the excess region defined by \citet{luhman2005} 
were $ 7.8 \pm 0.3$ per cent of the candidates to LMS,
$7.5 \pm 0.4$ per cent of the LMS confirmed as members and
$ 18.1 \pm 0.3$ per cent of the candidates to BD show IR excesses.
The right panel of Figure \ref{memberscc2} shows the IRAC
3.6$\mu$m-4.5$\mu$m vs. 5.8$\mu$m-8.0$\mu$m diagram showing 
the CTTS locus defined by \citet{hartmann2005a} and \citet{luhman2005},
were $12.0 \pm 0.4$ per cent of the LMS candidates,
$ 9.4 \pm 0.8$ per cent of the LMS confirmed as members and
$ 39.2 \pm 0.4$ per cent of the BD candidates show IR excesses.
\par
Using the Ks-8$\mu$m vs. Ks-24$\mu$m and the Ks-5.8$\mu$m vs. Ks-24$\mu$m
diagrams based on IRAC and MIPS photometry, \citet{luhman2010} show the distinction 
between circumstellar discs at different evolutionary stages. We do not have MIPS 
observations for the candidates and members considered in this work but we used
the Ks-8$\mu$m and Ks-5.8$\mu$m colour intervals defined by \citet{luhman2010} in order
to separate objects showing photospheric emissions from those showing primordial
discs and those showing transitional or evolved discs. Following this procedure
we obtain an estimation of the number fraction of such populations.
The left panel of Figure \ref{memberscc3} shows the Ks-5.8$\mu$m vs. Ks-8.0$\mu$m 
diagram, where we consider objects with excesses to be those with Ks-5.8$\mu$m $>$ 1 and 
Ks-8.0$\mu$m $>$ 1, resulting in a number fraction of $7.9 \pm 0.5$ per cent of the 
LMS candidates, $5.7 \pm 0.4 $ per cent for confirmed LMS and $\sim 47.0 \pm 0.5$ per cent 
for the photometric candidates to BD. In the transitional and evolved disc locus, 
defined as the region between $0.75 < Ks-8.0\mu m < 1.4$ and $0.3 < Ks-5.8\mu m < 1$,
we find 4 LMS members, 5 LMS candidates and 2 BD candidates for
number fractions of $7.4 \pm 0.4$ per cent, $2.3 \pm 0.5$ per cent and $15.6 \pm 0.5$
per cent respectively. The objects showing IR excesses indicative of
possible transitional or evolved discs are listed in  Table \ref{table:trans}. 
In the right panel of Figure \ref{memberscc3} 
we also show the excesses at longer wavelengths, using 
a Ks-4.6$\mu$m vs. Ks-12$\mu$m diagram from WISE data.
\begin{table*}
\centering
\begin{minipage}{400mm}
\caption{Members and candidates showing possible transitional or evolved discs.}
\begin{tiny}
\begin{tabular}{lccccccccccccccl}
\hline
$\alpha (J2000)$ & $\delta (J2000)$ & J & I & CH1 & CH2 & CH3 & CH4 & W1 & W2 & W3 & W4 & ST & WHa     & Av & Status \\
$[^\circ]$       & $[^\circ]$       &   &   &     &     &     &     &    &    &    &    &    & [$\AA$] & Av & Status \\
\hline
80.7117776526  & 1.51037395412  & 15.02  & 16.89 & 13.68 & 13.54 & 13.29 & 12.65 & 13.81 & 13.45 & 11.81  & 8.93  & $\cdots$  & $\cdots$  & $\cdots$  & Candidate \\
80.8808791321  & 2.22021102858  & 13.03  & 14.92 & 11.53 & 11.24 & 11.02 & 10.83 & 11.65 & 11.21 & 10.42  & 8.38  & $\cdots$  & $\cdots$  & $\cdots$  & Candidate \\
80.8829930224  & 2.17873892118  & 13.88  & 15.46 & 12.60 & 12.49 & 12.31 & 11.87 & 12.73 & 12.43 & 11.13  & 9.12  & $\cdots$  & $\cdots$  & $\cdots$  & Candidate \\
80.9343505087  & 1.69221474930  & 16.26  & 18.54 & 15.00 & 14.77 & 14.91 & 14.22 & 15.10 & 14.73 & 11.96  & 8.83  & M1.0      &  -3.5     & 6.1       & Member    \\
81.0206640554  & 2.15769232385  & 14.48  & 16.47 & 13.10 & 12.87 & 12.59 & 12.06 & 13.33 & 12.85 & 11.54  & 8.31  & $\cdots$  & $\cdots$  & $\cdots$  & Candidate \\
81.1102509066  & 1.84858069829  & 14.81  & 16.55 & 13.59 & 13.50 & 13.39 & 12.82 & 13.76 & 13.57 & 11.72  & 8.76  & M4.0      &  -2.1     & 1.1       & Member    \\
81.2452230354  & 1.42179289513  & 14.08  & 15.81 & 12.92 & 12.74 & 12.51 & 11.94 & 13.09 & 12.72 & 11.03  & 8.95  & M5.0      &  -7.4     & 0.0       & Member    \\
81.3513507400  & 1.68539771614  & 13.78  & 15.61 & 12.56 & 12.44 & 12.32 & 11.88 & 12.77 & 12.47 & 10.99  & 8.72  & M5.5      & -15.1     & 0.0       & Member    \\
81.4447905792  & 1.72510889268  & 13.15  & 14.54 & 11.97 & 11.90 & 11.72 & 11.08 & 12.10 & 11.84 &  9.91  & 6.74  & M3.5      & -10.1     & 0.0       & Member    \\
81.4683299453  & 2.36818676306  & 13.16  & 14.52 & 11.27 & 10.97 & 10.82 & 10.58 & 11.38 & 10.96 &  9.88  & 8.10  & $\cdots$  & $\cdots$  & $\cdots$  & Candidate \\
81.5062304737  & 1.54283333334  & 14.71  & 16.42 & 13.31 & 13.12 & 12.91 & 12.45 & 13.64 & 13.31 & 11.47  & 9.16  & $\cdots$  & $\cdots$  & $\cdots$  & Candidate \\
81.6041933360  & 1.87740776106  & 15.28  & 17.08 & 13.99 & 13.91 & 13.71 & 13.50 & 14.18 & 13.91 & 12.18  & 8.47  & M2.5      & $\cdots$  & 2.8       & Member    \\
81.7246496737  & 2.30153351910  & 15.00  & 16.67 & 13.76 & 13.55 & 13.36 & 12.60 & 13.86 & 13.52 & 12.31  & 8.62  & $\cdots$  & $\cdots$  & $\cdots$  & Candidate \\
81.8506189547  & 1.63383891153  & 17.09  & 20.32 & 15.18 & 15.04 & 15.34 & 14.32 & 15.66 & 15.44 & 12.41  & 9.14  & $\cdots$  & $\cdots$  & $\cdots$  & Candidate \\
82.3184071926  & 2.15333352098  & 13.76  & 15.46 & 12.40 & 12.16 & 11.84 & 11.32 & 12.52 & 12.12 & 10.92  & 8.91  & $\cdots$  & $\cdots$  & $\cdots$  & Candidate \\
82.4413576233  & 1.95882675222  & 14.02  & 15.73 & 12.72 & 12.53 & 12.24 & 11.69 & 12.86 & 12.46 & 11.04  & 8.71  & $\cdots$  & $\cdots$  & $\cdots$  & Candidate \\
82.5921100764  & 1.64194791673  & 15.15  & 17.32 & 13.52 & 13.31 & 13.05 & 12.92 & 13.71 & 13.30 & 12.30  & 8.59  & $\cdots$  & $\cdots$  & $\cdots$  & Candidate \\
\hline
\end{tabular}
\end{tiny}
\label{table:trans}
\end{minipage}
\end{table*}
\par
The samples of new confirmed members and candidates are
not affected by the photometric completeness of the IRAC observations
and the reported fractions are not biased towards the objects 
showing excesses. The number fraction of BD showing IR excesses turns out
to be systematically higher in all colour-colour diagrams suggesting 
that at an age of $\sim7$ Myr the disc fraction among BDs is larger than 
for LMS which could be an indicator of a slow evolution of discs around 
BDs. This supports previous results from \citet{luhman2012b} for the
Upper Scorpius association which with a estimated age of $\sim11$Myr 
\citep{pecaut2012} shows $\sim$25 per cent of objects later than M5 harbouring 
discs. Our results at a slight younger age support the scenario in
which discs around LMS and BD evolve in different time scales
in the sense that a higher fraction of BDs could retain their 
discs for longer periods of time.
\par
\begin{figure}
\includegraphics[width=80mm]{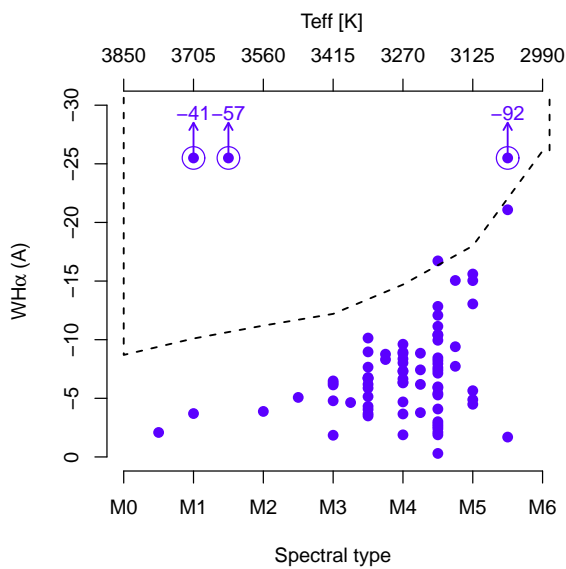}
\caption{Equivalent width of $H\alpha$ versus
spectral type and effective temperature (upper axis) for the
new confirmed LMS members.
Dotted lines indicate the limit between CTTS and WTTS
according to \citet{barradoynavascues2003}.
Circles indicates objects showing equivalent width of $H\alpha$
consistent with CTTS.
Arrows indicate objects with strong $H\alpha$
emission whose $WH\alpha$ is labelled.
We found 3.8 $\pm$ 0.5 per cent of the members of
25 Orionis group being CTTS (open circles).}
\label{fighast}
\end{figure}

Finally, low-resolution spectra provide an appropriate means to look for CTTS emission 
signatures such as H$\alpha$, the CaII 
triplet ($\lambda 8498$, $\lambda 8542$ 
and $\lambda 8662$) and HeI $\lambda 6676$.
The dividing line between CTTS and weak (non-accreting) T-Tauri stars
(WTTS) was first set in terms of the H$\alpha$ emission at W(H$\alpha$) $=-10$ {\AA} 
by \citet{herbig1988}. Afterwards, \citet{white2003} revised this 
classification and, based on high resolution optical spectra ($R\sim33000$), redefined
the classification considering both the H$\alpha$ equivalent width and 
the spectral type.
\citet{barradoynavascues2003} using low-resolution optical spectra 
developed an empirical classification method based on the H$\alpha$ 
equivalent width and the spectral type. 
Here we followed the criteria from \citet{barradoynavascues2003}
to classify the confirmed LMS as CTTS or WTTS.
Figure \ref{fighast} shows the resulting classification in the
$\rm W(H\alpha)$ vs. spectral-type plot. 
We obtained a number fraction of CTTS to WTTS of $3.8 \pm 0.5$ per cent
for objects earlier than M6. Our results are consistent
with those from the colour-colour diagrams and with
previous determinations by \citet{briceno2005}. On the other hand
the number fraction of CTTS presented here is larger than
the fraction reported by \citet{mcgehee2006}
($\sim 0.1$ per cent) that, as we explained in Section \ref{spatial},
most probably does not correspond to the 25 Orionis group.
%
%
%
%
%
%
\section{Summary and conclusions}\label{conclutions}
We have presented the first extensive study of the lower mass population 
($0.02 \le M/M_{\odot} \le 0.8$) of the 25 Ori group and its surrounding 
regions. We provide for this mass range the spatial distribution, mean age, 
IMF, and investigate the presence of circumstellar discs and accretion 
signatures. In the following we summarize our results:

\begin{itemize}

\item We have obtained a sample of 1246 photometric candidates to LMS and BDs 
in the 25 Ori group and its surroundings with estimated masses
within $0.02\lsim M/M_\odot\lsim0.8$, as well as 77 new LMS spectroscopically 
confirmed as members with spectral types in the range $\sim$M1 to $\sim$M5. 

\item We have characterised the possible contamination
of our photometric candidate sample using the Besan\c{c}on Galactic
model \citep{robin2003}. The expected mean contamination was found to be $\sim33$ 
field dwarf stars from the foreground per $deg^2$, which yields 
an expected average efficiency of $\sim82$ per cent for our photometric 
candidate selection. 

\item From the age distributions of the photometric
candidates and confirmed members we find mean ages of
$7.7\pm 0.6$ Myr and $6.1\pm 0.8$ Myr respectively.
These results are in good agreement with those
reported by \citet{briceno2005}.

\item We have derived the system-IMF for 25 Ori from a numerous
sample of photometric candidate LMS and BDs previously corrected
by the contamination from field stars. We have obtained
a system-IMF which can be well described by either a Kroupa power-law
function with indices $\alpha_3=-1.73\pm0.31$ and $\alpha_2=0.68\pm0.41$ in the mass
ranges $0.03<M/M_\odot<0.08$ and $0.08<M/M_\odot<0.5$ respectively,
or a Scalo log-normal function with coefficients $m_c=0.21^{+0.02}_{-0.02}$
and $\sigma=0.36\pm0.02$ in the mass range $0.03<M/M_\odot<0.8$.
For the log-normal representation of the IMF we obtained that the 
characteristic mass is consistent with an IMF independent of environmental 
properties, although the standard deviation is smaller than the system-IMF 
from \citet{chabrier2003a}. A sensitive photometric survey for the 
detection of a complete sample of BD down to $\sim0.01 M\odot$, and their 
spectroscopic counterparts is needed to confirm the slightly lower value 
of the standard deviation and whether this trend of the system-IMF is a 
consequence of a real deficit of BDs.

\item From the analysis of the spatial distribution of the photometric 
candidate sample we find the peak of the 25 Orionis overdensity
at $(\alpha,\delta)\sim(81\fdg2,1\fdg7)$ with a diameter of 
$\sim6.3$ pc. We have confirmed the East-West elongation of the 25 Orionis group,
also observed in the works from \citet{kharchenko2005} and \citet{briceno2005}, 
and rule out the southern extension proposed by \citet{mcgehee2006}. 
Additionally we found that the 25 Orionis group is projected over a filament
distributed along the south-west to north-east direction. We have also found 
that the spatial distributions of LMS and BDs in 25 Orionis are statistically 
indistinguishable. 

\item For the LMS we calculated the number fraction of stars showing IR 
excess in the J-H vs H-Ks diagram according to the CTTS locus from
\citet{meyer1997}. We found that $\sim 4$ per cent of the LMS show the levels 
of IR excess expected from CTTS with optically thick primordial discs.
We found similar fractions for the empirical locus defined in the 
$I-Ks$ vs $J-H$ colour-colour diagram where we find also that $\sim 12$ per cent 
of the BD show IR excesses indicative of primordial discs.

\item Using VISTA, IRAC and WISE photometry we have also found 
IR excess at longer wavelengths indicative of discs. 
Depending on the selected diagram we found that 
the number fraction of LMS showing IR excesses
is between $\sim$7 to $\sim$10 per cent while the number fraction 
for BDs increases up to $\sim$20 to $\sim$50 per cent.

\item Additionally we found that 11 members and candidates show
signs of possible transitional or evolved discs and that 
$3.8\pm0.5$ per cent of the LMS members show $H\alpha$ 
emission consistent with a CTTS nature, confirming the 
CTTS number fraction found by \citet{briceno2007a}. 

\item Our results show that the fraction of BD showing IR excesses
is higher than for LMS, supporting the scenario in which the
evolution of circumstellar discs around the least massive stars
could occur at larger time scales.

\end{itemize}

\section*{Acknowledgments}

This work has been supported in part by grant S1-200101144
and 200400829 from FONACIT, Venezuela. J. J. Downes, acknowledges 
support from CIDA, Venezuela and project 152160 from CONACyT, M\'exico.
C. Mateu acknowledges the support of the postdoctoral 
Fellowship of DGAPA-UNAM, M\'exico.
We thank Kevin Luhman for useful comments regarding the spectral
classification and Susan Tokarz, who is in charge of the
reduction and processing of Hectospec spectra.
\par
This work was based on observations collected at the
J\"urgen Stock 1-m Schmidt telescope of the National Observatory
of Llano del Hato Venezuela (NOV), which is operated by Centro
de Investigaciones de Astronom\'ia (CIDA) for the Ministerio del
Poder Popular para Ciencia y Tecnolog\'ia, Venezuela,
and at the MMT Observatory a joint facility of the Smithsonian
Institution and the University of Arizona.
This work is based [in part] on observations made with the Spitzer 
Space Telescope, which is operated by the Jet Propulsion Laboratory, 
California Institute of Technology under a contract with NASA. 
Support for this work was provided by NASA through an award issued
by JPL/Caltech.
This publication makes use of data products from the Wide-field Infrared Survey
Explorer, which is a joint project of the University of California, Los Angeles,
and the Jet Propulsion Laboratory/California Institute of Technology,
funded by the National Aeronautics and Space Administration.
We very much appreciate the great work done by the UK-based VISTA
consortium who built and commissioned the VISTA telescope and camera.
This work is based on observations made during VISTA science
verification, under the program ID 60.A-9285(B).
\par
We thank the assistance of the personnel, observers, telescope operators
and technical staff at CIDA and FLWO, who made possible the observations
at the J\"urgen Stock telescope of the Venezuela National Astronomical
Observatory (OAN) and at the MMT telescope at Fred Lawrence Whipple
Observatory (FLWO) of the Smithsonian Institution, especially
Perry Berlind, Daniel Cardozo, Orlando Contreras,
Franco Della Prugna, Emilio Falco, Freddy Moreno, Hern\'an Ram\'irez,
Carmen Rodr\'iguez, Richard Rojas, Gregore Rojas, Gerardo S\'anchez,
Gustavo S\'anchez and Ubaldo S\'anchez.
\par
This work makes an extensive use of R from the R Development Core
Team (2011) available at http://www.R-project.org/ and described
in \emph{R: A language and environment for statistical computing}
from R Foundation for Statistical Computing, Vienna, Austria,
ISBN 3-900051-07-0.
\par 
We thank the referee Phil Lucas for useful comments and
suggestions to the manuscript.



\bsp

\label{lastpage}

\end{document}